\documentstyle[aps]{revtex}

\begin{document}
\title{{\bf Anomalies Dismissed of Ambiguities and the Neutral Pion Decay}}
\author{O.A. Battistel* and G. Dallabona**}
\maketitle

\centerline{* Department. of Physics-CCNE, Universidade Federal de Santa
Maria}

\centerline{P.O. Box 5093, 97119-900, Santa Maria, RS, Brazil}

\centerline{** Department. of Physics-ICEx, Universidade Federal de Minas
Gerais}

\centerline{P.O. Box 702, 30161-970, Belo Horizonte, MG, Brazil}

\begin{abstract}
The correlation between the neutral electromagnetic pion decay, the
Sutherland-Veltman paradox and the $AVV$ triangle anomaly phenomenon is
discussed within the framework of an alternative strategy to handle the
divergences involved in the perturbative evaluation of the associated
physical amplitudes. We show that the general characteristic of the adopted
strategy allows us to recover the traditional treatment for the problem as
well as allows us to construct an alternative way to look at the problem
where the ambiguities play no relevant role.
\end{abstract}

\section{Introduction}

Our main intention in searching for the physical laws or formulating
theories and models is for the adequate and accurate description of the
available phenomenology. It does not make any sense to spend efforts in the
building of a theoretical apparatus if a well-characterized phenomenology
could not be put in accordance with the corresponding predictions. In the
establishment of the present knowledge about the fundamental particles and
interactions, many experimental data have been helpful. However, perhaps no
other data are so special as the neutral electromagnetic pion decay. Through
its phenomenology it was stated to be the most remarkable, subtle and
intriguing aspect of the Quantum Field Theory (QFT); the triangle anomalies 
\cite{Jackiw1,Jackiw2} whose theoretical implications go beyond the simple
accurate description of the experimental data. It is fair to say that our
present conception about the fundamental particles, two families of six
elements, the quarks and the leptons, is a direct implication of the neutral
electromagnetic pion decay phenomenology \cite{Livros}. A renormalizable
model is only achieved after the violations of the axial-vector Ward
identity of the $AVV$ (and $AAA$ ) one-loop triangle amplitude cancel, which
requires a specific set of constituent particles; precisely those of the
Standard Model \cite{Weinberg}. The triangle anomalies were mentioned in the
literature many years before their connection with the pion decay
phenomenology was estabilished, which was stated through the
Sutherland-Veltman paradox \cite{Sutherland-Veltman}. The standard methods
of current algebra, the $LSZ$ formalism and $PCAC$ hypothesis can be used to
show that if all three Ward identities are required to be satisfied, when
the $AVV$ amplitude is evaluated, the low-energy limit of such an amplitude
is not compatible with the predictions of a low-energy theorem which relates
the $AVV$ amplitude to the $\pi \rightarrow \gamma \gamma $ decay rate \cite
{Livros}. Only if an anomalous term is included in the $AVV$ calculated
amplitude, the divergence of the axial current will correspond to an
amplitude with the correct low-energy limit. The fundamental nature of the
anomaly phenomenon resides in the fact that no calculational method can
avoid the occurrence of at least one symmetry relation violation to get the
correct low-energy behaviour, which is a necessary requirement in order to
put the theory in accordance with the phenomenology. The last sentence
states that we cannot have a renormalizable theory with only one type of
fermion due to the fact that if the theory is in agreement with the
experimental data, it must violate a Ward identity, which eliminates the
chance for the renormalizability. Conversely, if the theory is
renormalizable, i.e., if all the Ward identities are preserved, the
corresponding theory should not be useful because it does not describe the
experimental data. All the general aspects previously mentioned about the
anomaly phenomenon are very well-known. What remains a problem is the
evaluation of the involved amplitudes and identification of the nature of
the violating terms in the perturbative calculations, i.e., the
justification of the anomalous term in the Ward identity connecting the $AVV$
triangle amplitude to the $PVV$ one, through the axial Ward identity. Such a
justification was proposed by Bell and Jackiw \cite{Jackiw1}, before the
introduction of the Dimensional Regularization (DR) \cite{DR}, which remains
even today as the accepted one. At least it can be found in almost all the
QFT textbooks \cite{Livros,Cheng-Li} in a closely related way to the point
of view of the reference \cite{Jackiw1}. The main point of the argumentation
resides in the divergent character, more precisely in the presence of
linearly divergent Feynman integrals in the external momenta contracted
expressions for the $AVV$ amplitude. In consequence, it is assumed that the
contracted expressions are ambiguous quantities since they are dependent on
the internal lines momentum routing, which means that two different choices
for the internal momentum labels lead to two different physical amplitudes.
An expression can be led into any other one by a shift on the integrating
momentum, but due to the linear divergence's degree, a surface term must be
added to the shifted amplitude, which will precisely constitute the
difference between them \cite{Livros,Cheng-Li}. The next step is to verify
that there are no possible choices for the internal lines momenta such that
all the Ward identities are simultaneously preserved. Given this
impossibility, the vector Ward identities are chosen to be preserved and the
axial one is assumed broken precisely by a quantity that corresponds to the
required anomalous term, which is necessary for the correct low-energy limit
of the $AVV$ physical amplitude. In spite of this point of view having well
succeeded in furnishment a justification for the anomalous term in the
perturbative evaluation of the involved amplitudes, some doubts remain about
the consistency of the arguments in a wider sense. In other words, if this
procedure is adopted to treat other physical problems, should we expect a
consistent description? This is due to the fact that from the point of view
of the DR, our best reference concerning the consistency in perturbative
calculations, all the momenta ambiguities are automatically eliminated. This
means that the ambiguities only play a relevant role in the triangle anomaly
justifications, where the DR cannot be applied. So, it is natural to ask
ourselves if the violations are a consequence of the non-existence of an
applicable consistent technique or if they represent a manifestation of a
fundamental and unavoidable phenomenon of the nature. If it is the case, and
it seems to be so, then we should expect that the source of the violations
must not be related to specific ingredients of the perturbative
calculations, due to the fact that if exact solutions were available, the
divergences, and consequently the ambiguities, would certainly be absent
while the anomalies should be still present in the problem. Another
important aspect is the existence of identities at the traces level, which
are valid before any calculations involving divergent structures have been
made, which relate the n-point Green's functions with an odd number of $%
\gamma _{5}$ Dirac matrix to those having an even number of such matrix.
This means that, even in the case where the DR cannot be applied to the
specific calculations involved, the mathematical structures to be evaluated
can be connected with other ones, through exact relations, which can be
evaluated within the scope of the DR. So, even in an indirect way, we can
test the consistency of a procedure using the results of the DR.

The purpose of the present work is to make some clarification to the points
mentioned above about the perturbative justifications of the triangle
anomalies. Our main tool for the investigation is the adoption of a very
general calculational strategy concerning the manipulations and calculations
of divergent amplitudes. Within this strategy it is possible to avoid the
explicit use of a regularization technique so that all the intrinsic
arbitrariness of the perturbative calculations remain still present in the
final expressions obtained this way. The mathematical objects which are
crucially dependent on the specific regularization philosophy eventually
applied are isolated from the independent ones. With this procedure it is
easily possible to map the results into those which may be obtained in any
traditional technique. This special property of our results will allow us to
extract interesting, clean and sound conclusions about the perturbative
justifications of the triangle anomalies. We will use these conclusions to
explain the neutral pion decay through the anomalous Ward identity \cite
{Jackiw1,Jackiw2,Livros} but without having recourse to internal momenta
ambiguities.

The work is organized in the following way. In the section II we introduce
the general aspects related to the discussion. In the section III we
consider the lowest order expression for the involved three-point functions
and their symmetry relations. The calculational strategy, to handle
divergences, is introduced in the section IV. The traditional way to look at
the problem is considered, from our results, in the section V. In the
section VI we consider the explicit evaluation of the three-point functions
involved and their symmetry relations verifications for, in the section VII,
present the final remarks and conclusions.

\section{The Neutral Pion Decay; Generalities and Sutherland-Veltman Paradox}

Let us introduce in this section the generalities involved in the
electromagnetic pion decay, which will be related to the discussion about
the $AVV$ anomaly. There are many textbooks where this issue can be found 
\cite{Livros}. We will follow in a closely related way the ref.\cite
{Cheng-Li} in order to state the problem. The electromagnetic decay of a
pion can be schematically represented by 
\begin{equation}
\pi \left( q\right) \rightarrow \gamma \left( p\right) +\gamma \left(
p^{\prime }\right) .
\end{equation}
The corresponding matrix element is given by 
\begin{equation}
\left\langle \gamma \left( p,\varepsilon _{1}\right) ,\gamma \left(
p^{\prime },\varepsilon _{2}\right) \mid \pi \left( q\right) \right\rangle
=i\left( 2\pi \right) ^{4}T_{\mu \nu }\left( q;p,p^{\prime }\right) \delta
^{4}\left( q-p-p^{\prime }\right) \varepsilon _{1}^{\mu }\left( p\right)
\varepsilon _{2}^{\nu }\left( p^{\prime }\right) .
\end{equation}
The tensor $T_{\mu \nu }$ is the amplitude connecting the external fields
and needs to be specified by a theory or model. Its general structure,
dictated by Lorentz invariance and CPT, can be written as 
\begin{equation}
T_{\mu \nu }\left( q;p,p^{\prime }\right) =\varepsilon _{\mu \nu \alpha
\beta }p_{\alpha }p_{\beta }^{\prime }\Gamma \left( q^{2}\right) .
\end{equation}
In terms of the electromagnetic current and the pion field we write 
\begin{equation}
T_{\mu \nu }\left( q;p,p^{\prime }\right) =e^{2}\int
d^{4}xd^{4}ye^{ipx+ip^{\prime }y}\left\langle 0\mid T\left( J_{\mu }\left(
x\right) J_{\nu }\left( y\right) \right) \mid \pi \left( q\right)
\right\rangle .
\end{equation}
The $PCAC$ hypothesis relates the pion field to the axial-vector current, $%
\partial ^{\lambda }A_{\lambda }^{a}=f_{\pi }m_{\pi }^{2}\pi ^{a}$, in such
a way that we can write $T_{\mu \nu }\left( q;p,p^{\prime }\right) $ only in
terms of currents. Using the $LSZ$ formalism it is possible to state such
relation as 
\begin{equation}
T_{\mu \nu }\left( q;p,p^{\prime }\right) =ie^{2}\frac{\left( q^{2}-m_{\pi
}^{2}\right) }{f_{\pi }m_{\pi }^{2}}\int d^{4}xd^{4}ye^{ipx+ip^{\prime
}y}\left\langle 0\mid T\left( \partial ^{\lambda }A_{\lambda }^{3}\left(
x\right) J_{\mu }\left( x\right) J_{\nu }\left( y\right) \right) \mid
0\right\rangle .
\end{equation}
The integral on the right hand side, on the other hand, can be related to
another amplitude by 
\begin{eqnarray}
q_{\lambda }T_{\lambda \mu \nu }\left( q;p,p^{\prime }\right) &=&\left(
-i\right) \int d^{4}xd^{4}ye^{ipx+ip^{\prime }y}\left\{ \left\langle 0\mid
T\left( \partial ^{\lambda }A_{\lambda }^{3}\left( x\right) J_{\mu }\left(
x\right) J_{\nu }\left( y\right) \right) \mid 0\right\rangle \right. 
\nonumber \\
&&+\left\langle 0\mid T\left( \left[ A_{0}^{3}\left( x\right) ,J_{\nu
}\left( y\right) \right] J_{\mu }\left( 0\right) \right) \mid 0\right\rangle
\delta \left( x_{0}-y_{0}\right)  \nonumber \\
&&+\left. \left\langle 0\mid T\left( \left[ A_{0}^{3}\left( x\right) ,J_{\mu
}\left( 0\right) \right] J_{\nu }\left( y\right) \right) \mid 0\right\rangle
\delta \left( x_{0}\right) \right\} ,
\end{eqnarray}
where 
\begin{equation}
T_{\lambda \mu \nu }\left( q;p,p^{\prime }\right) =\int
d^{4}xd^{4}ye^{ipx+ip^{\prime }y}\left\langle 0\mid T\left( A_{\lambda
}\left( x\right) J_{\nu }\left( y\right) J_{\mu }\left( 0\right) \right)
\mid 0\right\rangle .
\end{equation}

Now we can analyze what the relation (6) implies for the pion decay. The
first term can be associated with $T_{\mu \nu }$ amplitude which should
carry informations about the pion decay. The equation (6) states a
prediction about the value of the $\Gamma \left( q^{2}\right) $ in the
kinematical limit $q^{2}\rightarrow 0$. This is due to the fact that in the
limit $q_{\lambda }\rightarrow 0$, the left hand side $q_{\lambda
}T_{\lambda \mu \nu }$, must vanish. The two commutators on the right hand
side should also vanish by the current algebra properties. This means that $%
T_{\mu \nu }$ should also vanish in this limit which implies that $\Gamma
\left( q^{2}\right) \rightarrow 0$. Thus, if we identify $T_{\mu \nu }$ as
responsible for the pion decay description, and it obeys the predictions of
the above discussed low-energy limits, the decay rate $\pi ^{0}\rightarrow
2\gamma $ vanishes for $m_{\pi }^{2}\rightarrow 0$ and it is predicted a
very small value at the physical $m_{\pi }$. This is clearly in disagreement
with the experimental data \cite{Jackiw1,Livros}.

To arrive at this conclusion we can go by other track. We first note that
the $T_{\lambda \mu \nu }$ is a pseudo-tensor in such a way that its most
general structure, by Lorentz, CPT and Bose's statistics, is given by: 
\begin{eqnarray}
T^{\lambda \mu \nu } &=&\varepsilon ^{\mu \nu \omega \phi }p_{\omega
}p_{\phi }^{\prime }q^{\lambda }F_{1}\left( q^{2}\right) +\varepsilon
^{\lambda \mu \nu \omega }\left( p-p^{\prime }\right) _{\omega }\frac{q^{2}}{%
2}F_{3}\left( q^{2}\right)  \nonumber \\
&&+\left[ \varepsilon ^{\lambda \mu \omega \phi }p^{\prime \nu }-\varepsilon
^{\lambda \nu \omega \phi }p^{\mu }\right] p_{\omega }p_{\phi }^{\prime
}F_{2}\left( q^{2}\right)  \nonumber \\
&&+\left[ \varepsilon ^{\lambda \mu \omega \phi }p^{\nu }-\varepsilon
^{\lambda \nu \omega \phi }p^{\prime \mu }\right] p_{\omega }p_{\phi
}^{\prime }F_{3}\left( q^{2}\right) .
\end{eqnarray}
To obtain the above structure we have also required the conservation of the
vector current: $p_{\mu }T_{\lambda \mu \nu }=p_{\nu }^{\prime }T_{\lambda
\mu \nu }=0$ and put the external vectors on the mass shell $p^{2}=p^{\prime
2}=0$ such that $\left( p+p^{\prime }\right) ^{2}=q^{2}=2p\cdot p^{\prime }.$
The divergence of the axial current, or the contraction of the above
equation with $q_{\lambda }$, should give us the $T_{\mu \nu }$ amplitude,
which is 
\begin{equation}
q_{\lambda }T_{\lambda \mu \nu }=\varepsilon _{\mu \nu \alpha \beta
}p_{\alpha }p_{\beta }^{\prime }\left[ q^{2}\right] \left[ F_{1}+F_{3}\right]
.
\end{equation}
The left hand side vanishes in the limit $q_{\lambda }\rightarrow 0$, which
implies that the right hand side needs to vanish too. Again the conclusion
is that the ingredients used to state the result (8) lead to $\Gamma \left(
q^{2}\right) =0$ in the kinematical point $q^{2}=0$.

We can put what we have learned in these analyses as follows: The amplitude $%
T_{\lambda \mu \nu }$ can be calculated from the point of view of any theory
or model by using any calculational method. After this, four symmetry
properties can be verified: three Ward identities and a low-energy limit. If
all the Ward identities are obtained satisfied, then, necessarily, the
low-energy behavior for such a $T_{\lambda \mu \nu }$, which is related
through the axial Ward identity to the pion decay amplitude, will give us a
non-zero value, violating then the low-energy theorem obtained by the
current algebra methods. If, however, we want to obtain the low-energy limit
satisfied by constructing a $T_{\mu \nu }$ amplitude, which vanishes at $%
q^{2}=0$ and simultaneously satisfies all the Ward identities, then the
electromagnetic neutral pion decay rate is obtained in complete disagreement
with the experimental data. Therefore, if we want a theoretical tool which
adequately describes the phenomenology, it will be necessary to impose that
the correct $T_{\mu \nu }$ amplitude should have a non-zero low-energy
limit. As a consequence, if we want to construct the $T_{\lambda \mu \nu }$
such that in the limit $q_{\lambda }\rightarrow 0$%
\begin{equation}
\left\{ 
\begin{array}{l}
q_{\lambda }T_{\lambda \mu \nu }=0 \\ 
p_{\mu }T_{\lambda \mu \nu }=0 \\ 
p_{\nu }^{\prime }T_{\lambda \mu \nu }=0,
\end{array}
\right.
\end{equation}
we are forced to violate the axial Ward identity due to the fact that in
order to obtain a vanishing value for the right hand side of the axial Ward
identity, without to forbid the pion decay, it is necessary to assume the
presence of an extra term, i.e., $q_{\lambda }T_{\lambda \mu \nu }=T_{\mu
\nu }+C_{\mu \nu }$, that means to break the Ward identity by the amount $%
C_{\mu \nu }$ which is an anomalous term. The value for the anomalous term
is precisely the one of the low-energy value for the $T_{\mu \nu }$
amplitude which means that it is related to the experimental value for the
electromagnetic neutral pion decay rate.

The ingredients of the preceding discussion are well-known and constitute
the scope of the Sutherland-Veltman paradox \cite{Sutherland-Veltman}. What
remains as a question is the explicit evaluation of the involved amplitudes
in the context of perturbative calculations due to the presence of
divergences and, consequently, many types of arbitrariness. The main point
is: if an anomalous term, whose value is fixed by the pion decay
phenomenology, must be present in the calculated expressions, what is its
source? In other words, how can we justify the origin of the anomalous term
in perturbative calculations? Such a justification was given by Bell and
Jackiw \cite{Jackiw1}, that can be found in many QFT modern textbooks \cite
{Cheng-Li,Livros}, and it is deeply founded in the divergences aspects
involved, more precisely, in the linearly divergent character of the $%
T_{\lambda \mu \nu }$ perturbative amplitude. It was argued that there is an
intrinsic arbitrariness due to the fact that a shift in the integrating
momentum of the one-loop $T_{\lambda \mu \nu }$ amplitude gives raise to
another expression which differs from the original one by a surface term
whose coefficient is an undefined combination of the internal lines
momentum, which is arbitrary. The justification of the anomalous term is
then associated to the compulsory choices we must take, in order to give a
definite value for the $T_{\lambda \mu \nu }$ perturbative amplitude. In
spite of the adopted procedure being well-succeeded in the furnishment of a
justification for the anomalous term in the one-loop perturbative
calculations, some questions remained as interesting ones. First, the
justification is founded in the ambiguities which is an exclusive ingredient
of the perturbative calculation. However, the anomaly phenomenon is
predicted for the exact amplitudes, in which case the divergences and their
associated ambiguities are certainly absent. In view of these arguments we
should expect that the origin of the anomalous term does not reside in the
divergent aspects involved, which means that it cannot be associated to
ambiguities. The second aspect, that invites us to think about the
perturbative justification of triangle anomalies, is concerning the
consistency in perturbative calculations. We would like to look at all
physical amplitudes of all theories and models in the same way. This means
to interpret the divergences following an unique recipe. Our main tool to
handle divergences in QFT is undoubtedly the DR, which automatically
eliminates all the ambiguities even that the degree of divergence is higher
than the logarithmic one. This means that there is only one problem where
the ambiguities are called to play a relevant role; in the pseudo-amplitudes
where we find the triangle anomalies. So, the related question is: in a
general method to handle the divergences, which is capable to map all the
results of the DR, where this consistent technique can be applied, such that
no restrictions of applicability are present for the treatment of amplitudes
with an odd number of $\gamma _{5}$ matrix, can we explain the anomalies
through ambiguities? The reason for this question is immediate: if a method
maps the DR results it must automatically eliminate the ambiguities. So it
is expected, also through this line of reasoning, that the perturbative
justifications of the anomalous term cannot reside in the ambiguities.

After the preceding argumentation, we have stated our working lines; to
investigate in this context the origin of the triangle anomaly involved in
the pion decay phenomenology.

\section{$AVV$ and $PVV$ Amplitudes and their Symmetry Relations}

An explicit expression for $T_{\lambda \mu \nu }$ and $T_{\mu \nu }$, which
have appeared in the preceding section is only achieved after the
construction of a Lagrangian density or the involved currents. Given the
hadronic character of the pion, we need to look at the decay in terms of
intermediate states of quarks and anti-quarks. The lowest order Feynman
diagrams connecting the external fields are triangle diagrams, with only one
kind of quarks in the internal lines. Ignoring internal symmetries
operators, the currents involved are 
\begin{equation}
\left\{ 
\begin{array}{l}
V_{\mu }\left( x\right) =\overline{\Psi }\left( x\right) \gamma _{\mu }\Psi
\left( x\right) \\ 
A_{\mu }\left( x\right) =\overline{\Psi }\left( x\right) i\gamma _{\mu
}\gamma _{5}\Psi \left( x\right) \\ 
P\left( x\right) =\overline{\Psi }\left( x\right) \gamma _{5}\Psi \left(
x\right) ,
\end{array}
\right.
\end{equation}
where $\Psi \left( x\right) $ is the massive quark field obeying the Dirac
equation. As a consequence the above currents satisfy 
\begin{equation}
\left\{ 
\begin{array}{l}
\partial ^{\mu }V_{\mu }\left( x\right) =0 \\ 
\partial ^{\mu }A_{\mu }\left( x\right) =2miP\left( x\right) .
\end{array}
\right.
\end{equation}
The corresponding expressions $AVV$ and $PVV$ amplitudes in the lowest order
can be defined by \cite{Jackiw1,Cheng-Li,Livros} 
\begin{eqnarray}
\bullet T_{\lambda \mu \nu }^{AVV} &=&\int \frac{d^{4}k}{(2\pi )^{4}}%
Tr\left\{ i\gamma _{\lambda }\gamma _{5}\frac{1}{(\not{k}+{\not{k}}_{3})-m}%
\gamma _{\mu }\frac{1}{(\not{k}+{\not{k}}_{1})-m}\gamma _{\nu }\frac{1}{(%
\not{k}+{\not{k}}_{2})-m}\right\} \\
\bullet T_{\lambda \mu \nu }^{PVV} &=&\int \frac{d^{4}k}{(2\pi )^{4}}%
Tr\left\{ \gamma _{5}\frac{1}{(\not{k}+{\not{k}}_{3})-m}\gamma _{\mu }\frac{1%
}{(\not{k}+{\not{k}}_{1})-m}\gamma _{\nu }\frac{1}{(\not{k}+{\not{k}}_{2})-m}%
\right\} ,
\end{eqnarray}
where $k_{1},k_{2}$ and $k_{3}$ are the internal lines momenta, which are
related to the external ones by their differences as follows (see fig01) 
\begin{equation}
\left\{ 
\begin{array}{l}
k_{3}-k_{2}=q=p+p^{\prime } \\ 
k_{3}-k_{1}=p \\ 
k_{1}-k_{2}=p^{\prime }.
\end{array}
\right.
\end{equation}

\begin{center}
\setlength{\unitlength}{0.240900pt} 
\begin{picture}(5,5)(300,0)
\font\gnuplot=cmr10 at 10pt
\gnuplot
\put(70,260){$(a)$}
\put(90,345){\makebox(0,0)[l]{$i\gamma_\lambda \gamma_5$}}
\put(350,408){\makebox(0,0)[l]{$\gamma_\mu$}}
\put(350,272){\makebox(0,0)[l]{$\gamma_\nu$}}
\put(170,405){\makebox(0,0)[l]{$k+k_3$}}
\put(355,340){\makebox(0,0)[l]{$k+k_1$}}
\put(170,290){\makebox(0,0)[l]{$k+k_2$}}
\put(260,375){\vector(2,1){20}}
\put(280,305){\vector(-2,1){20}}
\put(341,350){\vector(0,-1){20}}
\put(325,400){$\bullet$}
\put(327,264){$\bullet$}
\put(188,333){$\bullet$}
\put(341,275){\line(0,1){140}}
\put(200,345){\line(2,1){140}}
\put(200,345){\line(2,-1){140}}
\end{picture}
\setlength{\unitlength}{0.240900pt} 
\begin{picture}(5,5)(-100,0)
\font\gnuplot=cmr10 at 10pt
\gnuplot
\put(70,260){$(b)$}
\put(140,345){\makebox(0,0)[l]{$\gamma_5$}}
\put(350,408){\makebox(0,0)[l]{$\gamma_\mu$}}
\put(350,272){\makebox(0,0)[l]{$\gamma_\nu$}}
\put(170,405){\makebox(0,0)[l]{$k+k_3$}}
\put(355,340){\makebox(0,0)[l]{$k+k_1$}}
\put(170,290){\makebox(0,0)[l]{$k+k_2$}}
\put(260,375){\vector(2,1){20}}
\put(280,305){\vector(-2,1){20}}
\put(341,350){\vector(0,-1){20}}
\put(325,400){$\bullet$}
\put(327,264){$\bullet$}
\put(188,333){$\bullet$}
\put(341,275){\line(0,1){140}}
\put(200,345){\line(2,1){140}}
\put(200,345){\line(2,-1){140}}
\end{picture}
\setlength{\unitlength}{0.240900pt} \ifx\plotpoint\undefined%
\newsavebox{\plotpoint}\fi
\sbox{\plotpoint}{\rule[-0.200pt]{0.400pt}{0.400pt}}%
\begin{picture}(750,450)(-450,-80)
\font\gnuplot=cmr10 at 10pt
\gnuplot
\put(150,180){$(c)$}
\put(313,256){\makebox(0,0)[l]{$\bullet$}}
\put(517,256){\makebox(0,0)[l]{$\bullet$}}
\put(425,204){\makebox(0,0)[l]{\vector(-1,0){20}}}
\put(425,327){\makebox(0,0)[l]{\vector(1,0){20}}}
\put(374,332){\makebox(0,0)[l]{$k+k_i$}}
\put(374,179){\makebox(0,0)[l]{$k+k_j$}}
\put(212,256){\makebox(0,0)[l]{$i\gamma_\mu \gamma_5$}}
\put(547,256){\makebox(0,0)[l]{$\gamma_\nu$}}
\multiput(323.59,256.00)(0.482,1.485){9}{\rule{0.116pt}{1.233pt}}
\multiput(322.17,256.00)(6.000,14.440){2}{\rule{0.400pt}{0.617pt}}
\multiput(329.00,273.59)(0.492,0.485){11}{\rule{0.500pt}{0.117pt}}
\multiput(329.00,272.17)(5.962,7.000){2}{\rule{0.250pt}{0.400pt}}
\multiput(336.00,280.59)(0.599,0.477){7}{\rule{0.580pt}{0.115pt}}
\multiput(336.00,279.17)(4.796,5.000){2}{\rule{0.290pt}{0.400pt}}
\multiput(342.00,285.60)(0.774,0.468){5}{\rule{0.700pt}{0.113pt}}
\multiput(342.00,284.17)(4.547,4.000){2}{\rule{0.350pt}{0.400pt}}
\multiput(348.00,289.61)(1.132,0.447){3}{\rule{0.900pt}{0.108pt}}
\multiput(348.00,288.17)(4.132,3.000){2}{\rule{0.450pt}{0.400pt}}
\multiput(354.00,292.61)(1.132,0.447){3}{\rule{0.900pt}{0.108pt}}
\multiput(354.00,291.17)(4.132,3.000){2}{\rule{0.450pt}{0.400pt}}
\put(360,295.17){\rule{1.300pt}{0.400pt}}
\multiput(360.00,294.17)(3.302,2.000){2}{\rule{0.650pt}{0.400pt}}
\put(366,297.17){\rule{1.500pt}{0.400pt}}
\multiput(366.00,296.17)(3.887,2.000){2}{\rule{0.750pt}{0.400pt}}
\put(373,299.17){\rule{1.300pt}{0.400pt}}
\multiput(373.00,298.17)(3.302,2.000){2}{\rule{0.650pt}{0.400pt}}
\put(379,301.17){\rule{1.300pt}{0.400pt}}
\multiput(379.00,300.17)(3.302,2.000){2}{\rule{0.650pt}{0.400pt}}
\put(385,302.67){\rule{1.445pt}{0.400pt}}
\multiput(385.00,302.17)(3.000,1.000){2}{\rule{0.723pt}{0.400pt}}
\put(391,303.67){\rule{1.445pt}{0.400pt}}
\multiput(391.00,303.17)(3.000,1.000){2}{\rule{0.723pt}{0.400pt}}
\put(397,304.67){\rule{1.445pt}{0.400pt}}
\multiput(397.00,304.17)(3.000,1.000){2}{\rule{0.723pt}{0.400pt}}
\put(416,305.67){\rule{1.445pt}{0.400pt}}
\multiput(416.00,305.17)(3.000,1.000){2}{\rule{0.723pt}{0.400pt}}
\put(403.0,306.0){\rule[-0.200pt]{3.132pt}{0.400pt}}
\put(428,305.67){\rule{1.445pt}{0.400pt}}
\multiput(428.00,306.17)(3.000,-1.000){2}{\rule{0.723pt}{0.400pt}}
\put(422.0,307.0){\rule[-0.200pt]{1.445pt}{0.400pt}}
\put(447,304.67){\rule{1.445pt}{0.400pt}}
\multiput(447.00,305.17)(3.000,-1.000){2}{\rule{0.723pt}{0.400pt}}
\put(453,303.67){\rule{1.445pt}{0.400pt}}
\multiput(453.00,304.17)(3.000,-1.000){2}{\rule{0.723pt}{0.400pt}}
\put(459,302.67){\rule{1.445pt}{0.400pt}}
\multiput(459.00,303.17)(3.000,-1.000){2}{\rule{0.723pt}{0.400pt}}
\put(465,301.17){\rule{1.300pt}{0.400pt}}
\multiput(465.00,302.17)(3.302,-2.000){2}{\rule{0.650pt}{0.400pt}}
\put(471,299.17){\rule{1.300pt}{0.400pt}}
\multiput(471.00,300.17)(3.302,-2.000){2}{\rule{0.650pt}{0.400pt}}
\put(477,297.17){\rule{1.500pt}{0.400pt}}
\multiput(477.00,298.17)(3.887,-2.000){2}{\rule{0.750pt}{0.400pt}}
\put(484,295.17){\rule{1.300pt}{0.400pt}}
\multiput(484.00,296.17)(3.302,-2.000){2}{\rule{0.650pt}{0.400pt}}
\multiput(490.00,293.95)(1.132,-0.447){3}{\rule{0.900pt}{0.108pt}}
\multiput(490.00,294.17)(4.132,-3.000){2}{\rule{0.450pt}{0.400pt}}
\multiput(496.00,290.95)(1.132,-0.447){3}{\rule{0.900pt}{0.108pt}}
\multiput(496.00,291.17)(4.132,-3.000){2}{\rule{0.450pt}{0.400pt}}
\multiput(502.00,287.94)(0.774,-0.468){5}{\rule{0.700pt}{0.113pt}}
\multiput(502.00,288.17)(4.547,-4.000){2}{\rule{0.350pt}{0.400pt}}
\multiput(508.00,283.93)(0.599,-0.477){7}{\rule{0.580pt}{0.115pt}}
\multiput(508.00,284.17)(4.796,-5.000){2}{\rule{0.290pt}{0.400pt}}
\multiput(514.00,278.93)(0.492,-0.485){11}{\rule{0.500pt}{0.117pt}}
\multiput(514.00,279.17)(5.962,-7.000){2}{\rule{0.250pt}{0.400pt}}
\multiput(521.59,267.88)(0.482,-1.485){9}{\rule{0.116pt}{1.233pt}}
\multiput(520.17,270.44)(6.000,-14.440){2}{\rule{0.400pt}{0.617pt}}
\put(434.0,306.0){\rule[-0.200pt]{3.132pt}{0.400pt}}
\put(323,256){\usebox{\plotpoint}}
\multiput(323.59,250.60)(0.482,-1.575){9}{\rule{0.116pt}{1.300pt}}
\multiput(322.17,253.30)(6.000,-15.302){2}{\rule{0.400pt}{0.650pt}}
\multiput(329.00,236.93)(0.492,-0.485){11}{\rule{0.500pt}{0.117pt}}
\multiput(329.00,237.17)(5.962,-7.000){2}{\rule{0.250pt}{0.400pt}}
\multiput(336.00,229.93)(0.599,-0.477){7}{\rule{0.580pt}{0.115pt}}
\multiput(336.00,230.17)(4.796,-5.000){2}{\rule{0.290pt}{0.400pt}}
\multiput(342.00,224.94)(0.774,-0.468){5}{\rule{0.700pt}{0.113pt}}
\multiput(342.00,225.17)(4.547,-4.000){2}{\rule{0.350pt}{0.400pt}}
\multiput(348.00,220.95)(1.132,-0.447){3}{\rule{0.900pt}{0.108pt}}
\multiput(348.00,221.17)(4.132,-3.000){2}{\rule{0.450pt}{0.400pt}}
\multiput(354.00,217.95)(1.132,-0.447){3}{\rule{0.900pt}{0.108pt}}
\multiput(354.00,218.17)(4.132,-3.000){2}{\rule{0.450pt}{0.400pt}}
\put(360,214.17){\rule{1.300pt}{0.400pt}}
\multiput(360.00,215.17)(3.302,-2.000){2}{\rule{0.650pt}{0.400pt}}
\put(366,212.17){\rule{1.500pt}{0.400pt}}
\multiput(366.00,213.17)(3.887,-2.000){2}{\rule{0.750pt}{0.400pt}}
\put(373,210.17){\rule{1.300pt}{0.400pt}}
\multiput(373.00,211.17)(3.302,-2.000){2}{\rule{0.650pt}{0.400pt}}
\put(379,208.17){\rule{1.300pt}{0.400pt}}
\multiput(379.00,209.17)(3.302,-2.000){2}{\rule{0.650pt}{0.400pt}}
\put(385,206.67){\rule{1.445pt}{0.400pt}}
\multiput(385.00,207.17)(3.000,-1.000){2}{\rule{0.723pt}{0.400pt}}
\put(391,205.67){\rule{1.445pt}{0.400pt}}
\multiput(391.00,206.17)(3.000,-1.000){2}{\rule{0.723pt}{0.400pt}}
\put(397,204.67){\rule{1.445pt}{0.400pt}}
\multiput(397.00,205.17)(3.000,-1.000){2}{\rule{0.723pt}{0.400pt}}
\put(416,203.67){\rule{1.445pt}{0.400pt}}
\multiput(416.00,204.17)(3.000,-1.000){2}{\rule{0.723pt}{0.400pt}}
\put(403.0,205.0){\rule[-0.200pt]{3.132pt}{0.400pt}}
\put(428,203.67){\rule{1.445pt}{0.400pt}}
\multiput(428.00,203.17)(3.000,1.000){2}{\rule{0.723pt}{0.400pt}}
\put(422.0,204.0){\rule[-0.200pt]{1.445pt}{0.400pt}}
\put(447,204.67){\rule{1.445pt}{0.400pt}}
\multiput(447.00,204.17)(3.000,1.000){2}{\rule{0.723pt}{0.400pt}}
\put(453,205.67){\rule{1.445pt}{0.400pt}}
\multiput(453.00,205.17)(3.000,1.000){2}{\rule{0.723pt}{0.400pt}}
\put(459,206.67){\rule{1.445pt}{0.400pt}}
\multiput(459.00,206.17)(3.000,1.000){2}{\rule{0.723pt}{0.400pt}}
\put(465,208.17){\rule{1.300pt}{0.400pt}}
\multiput(465.00,207.17)(3.302,2.000){2}{\rule{0.650pt}{0.400pt}}
\put(471,210.17){\rule{1.300pt}{0.400pt}}
\multiput(471.00,209.17)(3.302,2.000){2}{\rule{0.650pt}{0.400pt}}
\put(477,212.17){\rule{1.500pt}{0.400pt}}
\multiput(477.00,211.17)(3.887,2.000){2}{\rule{0.750pt}{0.400pt}}
\put(484,214.17){\rule{1.300pt}{0.400pt}}
\multiput(484.00,213.17)(3.302,2.000){2}{\rule{0.650pt}{0.400pt}}
\multiput(490.00,216.61)(1.132,0.447){3}{\rule{0.900pt}{0.108pt}}
\multiput(490.00,215.17)(4.132,3.000){2}{\rule{0.450pt}{0.400pt}}
\multiput(496.00,219.61)(1.132,0.447){3}{\rule{0.900pt}{0.108pt}}
\multiput(496.00,218.17)(4.132,3.000){2}{\rule{0.450pt}{0.400pt}}
\multiput(502.00,222.60)(0.774,0.468){5}{\rule{0.700pt}{0.113pt}}
\multiput(502.00,221.17)(4.547,4.000){2}{\rule{0.350pt}{0.400pt}}
\multiput(508.00,226.59)(0.599,0.477){7}{\rule{0.580pt}{0.115pt}}
\multiput(508.00,225.17)(4.796,5.000){2}{\rule{0.290pt}{0.400pt}}
\multiput(514.00,231.59)(0.492,0.485){11}{\rule{0.500pt}{0.117pt}}
\multiput(514.00,230.17)(5.962,7.000){2}{\rule{0.250pt}{0.400pt}}
\multiput(521.59,238.00)(0.482,1.575){9}{\rule{0.116pt}{1.300pt}}
\multiput(520.17,238.00)(6.000,15.302){2}{\rule{0.400pt}{0.650pt}}
\put(434.0,205.0){\rule[-0.200pt]{3.132pt}{0.400pt}}
\end{picture}
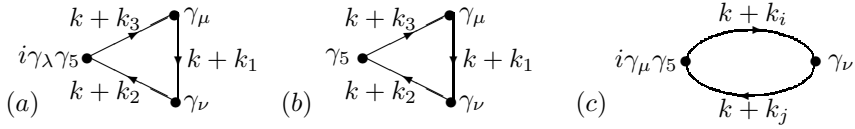
\vskip-1.8cm {\it Fig01: Diagrammatic representation for the }${\it AVV}$%
{\it \ and }${\it PVV}${\it \ three-point functions and for the $AV$
two-point function, figs.(a), (b),and (c), respectively. }
\end{center}

The necessary crossed diagram, to construct the physical process, can be
obtained by changing \ $\mu $ and $\nu $ and changing the internal lines
arbitrary momenta $k^{\prime }s$ to $l^{\prime }s$, satisfying then 
\begin{equation}
\left\{ 
\begin{array}{l}
l_{3}-l_{2}=q=p+p^{\prime } \\ 
l_{3}-l_{1}=p^{\prime } \\ 
l_{1}-l_{2}=p.
\end{array}
\right.
\end{equation}
The explicit evaluation of the above structures should allow us to verify
what we have announced in the context of the Sutherland-Veltman paradox.
This is precisely the step that generates the difficulties we have in
understanding the involved ingredients. Both structures are, by power
counting, superficially linearly divergent. In practice, after the traces
evaluation, $T_{\mu \nu }^{PVV}$ exhibits a finite character while $%
T_{\lambda \mu \nu }^{AVV}$ remains linearly divergent. The last sentence
implies that, in order to verify the four symmetry properties previously
discussed, we need to handle, in some way, divergent integrals. Such \
calculations involve, as it is well-known, many types of arbitrariness.
Consequently only the adoption of a consistent strategy to handle the
divergences can avoid the transformation of the arbitrariness into
ambiguities. By ambiguities, we mean all the dependence on the final results
of the specific choices required by the involved arbitrariness. The worse
aspect related to such ambiguities is the invariable breaking of some
symmetries associated to them. In what follows, the aspects arbitrariness,
ambiguities and symmetry violations, will play the most important role in
our discussions. Before the explicit evaluation, let us verify what we can
expect to find from general grounds for the amplitudes.

The three-point functions $T_{\mu \nu }^{PVV}$ and $T_{\lambda \mu \nu
}^{AVV}$ should exhibit the symmetry properties of the amplitudes $T_{\mu
\nu }$ and $T_{\lambda \mu \nu }$ which we have presented in the section II.
The verification of such properties is made by taking the contractions of
the Green's functions with the external momenta associated with the
respective Lorentz index. Such operation can be used in the level of traces
to state conditions to be fulfilled in terms of the Green's functions with a
lower number of points. In principle, a divergent Green's function with a
lower number of points presents a higher degree of divergence. This means,
for example, that by successive contractions we can associate a logarithmic
divergent Green's function with those linear and quadratic ones. This type
of association may allow us to test the choices we need to make, in the
calculations involving divergent structures, in a broader sense. For our
present problem, this means to generate relations between the one-time
contracted three-point functions with other three and two-point functions.

To construct such relations we may consider identities like \cite
{Livros,Jackiw2} 
\begin{eqnarray}
&&(k_{3}-k_{2})_{\lambda }\left\{ \gamma _{\nu }\frac{1}{(\not{k}+{\not{k}}%
_{2})-m}i\gamma _{\lambda }\gamma _{5}\frac{1}{(\not{k}+{\not{k}}_{3})-m}%
\gamma _{\mu }\frac{1}{(\not{k}+{\not{k}}_{1})-m}\right\} =-\left\{ i\gamma
_{\nu }\gamma _{5}\frac{1}{(\not{k}+{\not{k}}_{3})-m}\gamma _{\mu }\frac{1}{(%
\not{k}+{\not{k}}_{1})-m}\right\}  \nonumber \\
&&+\left\{ \gamma _{\nu }\frac{1}{(\not{k}+{\not{k}}_{2})-m}i\gamma _{\mu
}\gamma _{5}\frac{1}{(\not{k}+{\not{k}}_{1})-m}\right\} -2mi\left\{ \gamma
_{\nu }\frac{1}{(\not{k}+{\not{k}}_{2})-m}\gamma _{5}\frac{1}{(\not{k}+{\not{%
k}}_{3})-m}\gamma _{\mu }\frac{1}{(\not{k}+{\not{k}}_{1})-m}\right\} .
\end{eqnarray}
Taking the traces and integrating in the momentum $k$ in both sides, this
means that (see fig02) 
\begin{equation}
\left( k_{3}-k_{2}\right) _{\lambda }T_{\lambda \mu \nu }^{AVV}=-2miT_{\mu
\nu }^{PVV}+T_{\mu \nu }^{AV}\left( k_{1},k_{2};m\right) -T_{\nu \mu
}^{AV}\left( k_{3},k_{1};m\right) .
\end{equation}

\begin{center}
\setlength{\unitlength}{0.240900pt} 
\begin{picture}(5,5)(100,0)
\font\gnuplot=cmr10 at 10pt
\gnuplot
\put(480,235){\line(-1,0){10}}
\put(480,445){\line(-1,0){10}}
\put(480,235){\line(0,1){210}}
\put(80,235){\line(1,0){10}}
\put(80,445){\line(1,0){10}}
\put(80,235){\line(0,1){210}}
\put(70,345){\makebox(0,0)[r]{$(k_2-k_3)^\lambda$}}
\put(90,345){\makebox(0,0)[l]{$i\gamma_\lambda \gamma_5$}}
\put(350,408){\makebox(0,0)[l]{$\gamma_\mu$}}
\put(350,272){\makebox(0,0)[l]{$\gamma_\nu$}}
\put(170,405){\makebox(0,0)[l]{$k+k_3$}}
\put(355,340){\makebox(0,0)[l]{$k+k_1$}}
\put(170,290){\makebox(0,0)[l]{$k+k_2$}}
\put(260,375){\vector(2,1){20}}
\put(280,305){\vector(-2,1){20}}
\put(341,350){\vector(0,-1){20}}
\put(325,400){$\bullet$}
\put(327,264){$\bullet$}
\put(188,333){$\bullet$}
\put(341,275){\line(0,1){140}}
\put(200,345){\line(2,1){140}}
\put(200,345){\line(2,-1){140}}
\put(510,330){$=$}
\end{picture}
\setlength{\unitlength}{0.240900pt} 
\begin{picture}(5,5)(-380,0)
\font\gnuplot=cmr10 at 10pt
\gnuplot
\put(50,345){\makebox(0,0)[l]{$2im$}}
\put(130,235){\line(1,0){10}}
\put(130,445){\line(1,0){10}}
\put(130,235){\line(0,1){210}}
\put(480,235){\line(-1,0){10}}
\put(480,445){\line(-1,0){10}}
\put(480,235){\line(0,1){210}}
\put(140,345){\makebox(0,0)[l]{$\gamma_5$}}
\put(350,408){\makebox(0,0)[l]{$\gamma_\mu$}}
\put(350,272){\makebox(0,0)[l]{$\gamma_\nu$}}
\put(170,405){\makebox(0,0)[l]{$k+k_3$}}
\put(355,340){\makebox(0,0)[l]{$k+k_1$}}
\put(170,290){\makebox(0,0)[l]{$k+k_2$}}
\put(260,375){\vector(2,1){20}}
\put(280,305){\vector(-2,1){20}}
\put(341,350){\vector(0,-1){20}}
\put(325,400){$\bullet$}
\put(327,264){$\bullet$}
\put(188,333){$\bullet$}
\put(341,275){\line(0,1){140}}
\put(200,345){\line(2,1){140}}
\put(200,345){\line(2,-1){140}}
\put(500,330){$-$}
\end{picture}
\setlength{\unitlength}{0.240900pt} \ifx\plotpoint\undefined%
\newsavebox{\plotpoint}\fi
\sbox{\plotpoint}{\rule[-0.200pt]{0.400pt}{0.400pt}}%
\begin{picture}(750,450)(-690,-80)
\font\gnuplot=cmr10 at 10pt
\gnuplot
\put(313,256){\makebox(0,0)[l]{$\bullet$}}
\put(517,256){\makebox(0,0)[l]{$\bullet$}}
\put(425,204){\makebox(0,0)[l]{\vector(-1,0){20}}}
\put(425,327){\makebox(0,0)[l]{\vector(1,0){20}}}
\put(374,332){\makebox(0,0)[l]{$k+k_1$}}
\put(374,179){\makebox(0,0)[l]{$k+k_2$}}
\put(212,256){\makebox(0,0)[l]{$i\gamma_\mu \gamma_5$}}
\put(547,256){\makebox(0,0)[l]{$\gamma_\nu$}}
\multiput(323.59,256.00)(0.482,1.485){9}{\rule{0.116pt}{1.233pt}}
\multiput(322.17,256.00)(6.000,14.440){2}{\rule{0.400pt}{0.617pt}}
\multiput(329.00,273.59)(0.492,0.485){11}{\rule{0.500pt}{0.117pt}}
\multiput(329.00,272.17)(5.962,7.000){2}{\rule{0.250pt}{0.400pt}}
\multiput(336.00,280.59)(0.599,0.477){7}{\rule{0.580pt}{0.115pt}}
\multiput(336.00,279.17)(4.796,5.000){2}{\rule{0.290pt}{0.400pt}}
\multiput(342.00,285.60)(0.774,0.468){5}{\rule{0.700pt}{0.113pt}}
\multiput(342.00,284.17)(4.547,4.000){2}{\rule{0.350pt}{0.400pt}}
\multiput(348.00,289.61)(1.132,0.447){3}{\rule{0.900pt}{0.108pt}}
\multiput(348.00,288.17)(4.132,3.000){2}{\rule{0.450pt}{0.400pt}}
\multiput(354.00,292.61)(1.132,0.447){3}{\rule{0.900pt}{0.108pt}}
\multiput(354.00,291.17)(4.132,3.000){2}{\rule{0.450pt}{0.400pt}}
\put(360,295.17){\rule{1.300pt}{0.400pt}}
\multiput(360.00,294.17)(3.302,2.000){2}{\rule{0.650pt}{0.400pt}}
\put(366,297.17){\rule{1.500pt}{0.400pt}}
\multiput(366.00,296.17)(3.887,2.000){2}{\rule{0.750pt}{0.400pt}}
\put(373,299.17){\rule{1.300pt}{0.400pt}}
\multiput(373.00,298.17)(3.302,2.000){2}{\rule{0.650pt}{0.400pt}}
\put(379,301.17){\rule{1.300pt}{0.400pt}}
\multiput(379.00,300.17)(3.302,2.000){2}{\rule{0.650pt}{0.400pt}}
\put(385,302.67){\rule{1.445pt}{0.400pt}}
\multiput(385.00,302.17)(3.000,1.000){2}{\rule{0.723pt}{0.400pt}}
\put(391,303.67){\rule{1.445pt}{0.400pt}}
\multiput(391.00,303.17)(3.000,1.000){2}{\rule{0.723pt}{0.400pt}}
\put(397,304.67){\rule{1.445pt}{0.400pt}}
\multiput(397.00,304.17)(3.000,1.000){2}{\rule{0.723pt}{0.400pt}}
\put(416,305.67){\rule{1.445pt}{0.400pt}}
\multiput(416.00,305.17)(3.000,1.000){2}{\rule{0.723pt}{0.400pt}}
\put(403.0,306.0){\rule[-0.200pt]{3.132pt}{0.400pt}}
\put(428,305.67){\rule{1.445pt}{0.400pt}}
\multiput(428.00,306.17)(3.000,-1.000){2}{\rule{0.723pt}{0.400pt}}
\put(422.0,307.0){\rule[-0.200pt]{1.445pt}{0.400pt}}
\put(447,304.67){\rule{1.445pt}{0.400pt}}
\multiput(447.00,305.17)(3.000,-1.000){2}{\rule{0.723pt}{0.400pt}}
\put(453,303.67){\rule{1.445pt}{0.400pt}}
\multiput(453.00,304.17)(3.000,-1.000){2}{\rule{0.723pt}{0.400pt}}
\put(459,302.67){\rule{1.445pt}{0.400pt}}
\multiput(459.00,303.17)(3.000,-1.000){2}{\rule{0.723pt}{0.400pt}}
\put(465,301.17){\rule{1.300pt}{0.400pt}}
\multiput(465.00,302.17)(3.302,-2.000){2}{\rule{0.650pt}{0.400pt}}
\put(471,299.17){\rule{1.300pt}{0.400pt}}
\multiput(471.00,300.17)(3.302,-2.000){2}{\rule{0.650pt}{0.400pt}}
\put(477,297.17){\rule{1.500pt}{0.400pt}}
\multiput(477.00,298.17)(3.887,-2.000){2}{\rule{0.750pt}{0.400pt}}
\put(484,295.17){\rule{1.300pt}{0.400pt}}
\multiput(484.00,296.17)(3.302,-2.000){2}{\rule{0.650pt}{0.400pt}}
\multiput(490.00,293.95)(1.132,-0.447){3}{\rule{0.900pt}{0.108pt}}
\multiput(490.00,294.17)(4.132,-3.000){2}{\rule{0.450pt}{0.400pt}}
\multiput(496.00,290.95)(1.132,-0.447){3}{\rule{0.900pt}{0.108pt}}
\multiput(496.00,291.17)(4.132,-3.000){2}{\rule{0.450pt}{0.400pt}}
\multiput(502.00,287.94)(0.774,-0.468){5}{\rule{0.700pt}{0.113pt}}
\multiput(502.00,288.17)(4.547,-4.000){2}{\rule{0.350pt}{0.400pt}}
\multiput(508.00,283.93)(0.599,-0.477){7}{\rule{0.580pt}{0.115pt}}
\multiput(508.00,284.17)(4.796,-5.000){2}{\rule{0.290pt}{0.400pt}}
\multiput(514.00,278.93)(0.492,-0.485){11}{\rule{0.500pt}{0.117pt}}
\multiput(514.00,279.17)(5.962,-7.000){2}{\rule{0.250pt}{0.400pt}}
\multiput(521.59,267.88)(0.482,-1.485){9}{\rule{0.116pt}{1.233pt}}
\multiput(520.17,270.44)(6.000,-14.440){2}{\rule{0.400pt}{0.617pt}}
\put(434.0,306.0){\rule[-0.200pt]{3.132pt}{0.400pt}}
\put(323,256){\usebox{\plotpoint}}
\multiput(323.59,250.60)(0.482,-1.575){9}{\rule{0.116pt}{1.300pt}}
\multiput(322.17,253.30)(6.000,-15.302){2}{\rule{0.400pt}{0.650pt}}
\multiput(329.00,236.93)(0.492,-0.485){11}{\rule{0.500pt}{0.117pt}}
\multiput(329.00,237.17)(5.962,-7.000){2}{\rule{0.250pt}{0.400pt}}
\multiput(336.00,229.93)(0.599,-0.477){7}{\rule{0.580pt}{0.115pt}}
\multiput(336.00,230.17)(4.796,-5.000){2}{\rule{0.290pt}{0.400pt}}
\multiput(342.00,224.94)(0.774,-0.468){5}{\rule{0.700pt}{0.113pt}}
\multiput(342.00,225.17)(4.547,-4.000){2}{\rule{0.350pt}{0.400pt}}
\multiput(348.00,220.95)(1.132,-0.447){3}{\rule{0.900pt}{0.108pt}}
\multiput(348.00,221.17)(4.132,-3.000){2}{\rule{0.450pt}{0.400pt}}
\multiput(354.00,217.95)(1.132,-0.447){3}{\rule{0.900pt}{0.108pt}}
\multiput(354.00,218.17)(4.132,-3.000){2}{\rule{0.450pt}{0.400pt}}
\put(360,214.17){\rule{1.300pt}{0.400pt}}
\multiput(360.00,215.17)(3.302,-2.000){2}{\rule{0.650pt}{0.400pt}}
\put(366,212.17){\rule{1.500pt}{0.400pt}}
\multiput(366.00,213.17)(3.887,-2.000){2}{\rule{0.750pt}{0.400pt}}
\put(373,210.17){\rule{1.300pt}{0.400pt}}
\multiput(373.00,211.17)(3.302,-2.000){2}{\rule{0.650pt}{0.400pt}}
\put(379,208.17){\rule{1.300pt}{0.400pt}}
\multiput(379.00,209.17)(3.302,-2.000){2}{\rule{0.650pt}{0.400pt}}
\put(385,206.67){\rule{1.445pt}{0.400pt}}
\multiput(385.00,207.17)(3.000,-1.000){2}{\rule{0.723pt}{0.400pt}}
\put(391,205.67){\rule{1.445pt}{0.400pt}}
\multiput(391.00,206.17)(3.000,-1.000){2}{\rule{0.723pt}{0.400pt}}
\put(397,204.67){\rule{1.445pt}{0.400pt}}
\multiput(397.00,205.17)(3.000,-1.000){2}{\rule{0.723pt}{0.400pt}}
\put(416,203.67){\rule{1.445pt}{0.400pt}}
\multiput(416.00,204.17)(3.000,-1.000){2}{\rule{0.723pt}{0.400pt}}
\put(403.0,205.0){\rule[-0.200pt]{3.132pt}{0.400pt}}
\put(428,203.67){\rule{1.445pt}{0.400pt}}
\multiput(428.00,203.17)(3.000,1.000){2}{\rule{0.723pt}{0.400pt}}
\put(422.0,204.0){\rule[-0.200pt]{1.445pt}{0.400pt}}
\put(447,204.67){\rule{1.445pt}{0.400pt}}
\multiput(447.00,204.17)(3.000,1.000){2}{\rule{0.723pt}{0.400pt}}
\put(453,205.67){\rule{1.445pt}{0.400pt}}
\multiput(453.00,205.17)(3.000,1.000){2}{\rule{0.723pt}{0.400pt}}
\put(459,206.67){\rule{1.445pt}{0.400pt}}
\multiput(459.00,206.17)(3.000,1.000){2}{\rule{0.723pt}{0.400pt}}
\put(465,208.17){\rule{1.300pt}{0.400pt}}
\multiput(465.00,207.17)(3.302,2.000){2}{\rule{0.650pt}{0.400pt}}
\put(471,210.17){\rule{1.300pt}{0.400pt}}
\multiput(471.00,209.17)(3.302,2.000){2}{\rule{0.650pt}{0.400pt}}
\put(477,212.17){\rule{1.500pt}{0.400pt}}
\multiput(477.00,211.17)(3.887,2.000){2}{\rule{0.750pt}{0.400pt}}
\put(484,214.17){\rule{1.300pt}{0.400pt}}
\multiput(484.00,213.17)(3.302,2.000){2}{\rule{0.650pt}{0.400pt}}
\multiput(490.00,216.61)(1.132,0.447){3}{\rule{0.900pt}{0.108pt}}
\multiput(490.00,215.17)(4.132,3.000){2}{\rule{0.450pt}{0.400pt}}
\multiput(496.00,219.61)(1.132,0.447){3}{\rule{0.900pt}{0.108pt}}
\multiput(496.00,218.17)(4.132,3.000){2}{\rule{0.450pt}{0.400pt}}
\multiput(502.00,222.60)(0.774,0.468){5}{\rule{0.700pt}{0.113pt}}
\multiput(502.00,221.17)(4.547,4.000){2}{\rule{0.350pt}{0.400pt}}
\multiput(508.00,226.59)(0.599,0.477){7}{\rule{0.580pt}{0.115pt}}
\multiput(508.00,225.17)(4.796,5.000){2}{\rule{0.290pt}{0.400pt}}
\multiput(514.00,231.59)(0.492,0.485){11}{\rule{0.500pt}{0.117pt}}
\multiput(514.00,230.17)(5.962,7.000){2}{\rule{0.250pt}{0.400pt}}
\multiput(521.59,238.00)(0.482,1.575){9}{\rule{0.116pt}{1.300pt}}
\multiput(520.17,238.00)(6.000,15.302){2}{\rule{0.400pt}{0.650pt}}
\put(434.0,205.0){\rule[-0.200pt]{3.132pt}{0.400pt}}
\put(607,255){\makebox(0,0)[l]{$+$}}
\end{picture}
\setlength{\unitlength}{0.240900pt} \ifx\plotpoint\undefined%
\newsavebox{\plotpoint}\fi
\sbox{\plotpoint}{\rule[-0.200pt]{0.400pt}{0.400pt}}%
\begin{picture}(750,450)(-360,-80)
\font\gnuplot=cmr10 at 10pt
\gnuplot
\put(313,256){\makebox(0,0)[l]{$\bullet$}}
\put(517,256){\makebox(0,0)[l]{$\bullet$}}
\put(425,204){\makebox(0,0)[l]{\vector(-1,0){20}}}
\put(425,327){\makebox(0,0)[l]{\vector(1,0){20}}}
\put(374,332){\makebox(0,0)[l]{$k+k_3$}}
\put(374,179){\makebox(0,0)[l]{$k+k_1$}}
\put(212,256){\makebox(0,0)[l]{$i\gamma_\nu \gamma_5$}}
\put(547,256){\makebox(0,0)[l]{$\gamma_\mu$}}
\multiput(323.59,256.00)(0.482,1.485){9}{\rule{0.116pt}{1.233pt}}
\multiput(322.17,256.00)(6.000,14.440){2}{\rule{0.400pt}{0.617pt}}
\multiput(329.00,273.59)(0.492,0.485){11}{\rule{0.500pt}{0.117pt}}
\multiput(329.00,272.17)(5.962,7.000){2}{\rule{0.250pt}{0.400pt}}
\multiput(336.00,280.59)(0.599,0.477){7}{\rule{0.580pt}{0.115pt}}
\multiput(336.00,279.17)(4.796,5.000){2}{\rule{0.290pt}{0.400pt}}
\multiput(342.00,285.60)(0.774,0.468){5}{\rule{0.700pt}{0.113pt}}
\multiput(342.00,284.17)(4.547,4.000){2}{\rule{0.350pt}{0.400pt}}
\multiput(348.00,289.61)(1.132,0.447){3}{\rule{0.900pt}{0.108pt}}
\multiput(348.00,288.17)(4.132,3.000){2}{\rule{0.450pt}{0.400pt}}
\multiput(354.00,292.61)(1.132,0.447){3}{\rule{0.900pt}{0.108pt}}
\multiput(354.00,291.17)(4.132,3.000){2}{\rule{0.450pt}{0.400pt}}
\put(360,295.17){\rule{1.300pt}{0.400pt}}
\multiput(360.00,294.17)(3.302,2.000){2}{\rule{0.650pt}{0.400pt}}
\put(366,297.17){\rule{1.500pt}{0.400pt}}
\multiput(366.00,296.17)(3.887,2.000){2}{\rule{0.750pt}{0.400pt}}
\put(373,299.17){\rule{1.300pt}{0.400pt}}
\multiput(373.00,298.17)(3.302,2.000){2}{\rule{0.650pt}{0.400pt}}
\put(379,301.17){\rule{1.300pt}{0.400pt}}
\multiput(379.00,300.17)(3.302,2.000){2}{\rule{0.650pt}{0.400pt}}
\put(385,302.67){\rule{1.445pt}{0.400pt}}
\multiput(385.00,302.17)(3.000,1.000){2}{\rule{0.723pt}{0.400pt}}
\put(391,303.67){\rule{1.445pt}{0.400pt}}
\multiput(391.00,303.17)(3.000,1.000){2}{\rule{0.723pt}{0.400pt}}
\put(397,304.67){\rule{1.445pt}{0.400pt}}
\multiput(397.00,304.17)(3.000,1.000){2}{\rule{0.723pt}{0.400pt}}
\put(416,305.67){\rule{1.445pt}{0.400pt}}
\multiput(416.00,305.17)(3.000,1.000){2}{\rule{0.723pt}{0.400pt}}
\put(403.0,306.0){\rule[-0.200pt]{3.132pt}{0.400pt}}
\put(428,305.67){\rule{1.445pt}{0.400pt}}
\multiput(428.00,306.17)(3.000,-1.000){2}{\rule{0.723pt}{0.400pt}}
\put(422.0,307.0){\rule[-0.200pt]{1.445pt}{0.400pt}}
\put(447,304.67){\rule{1.445pt}{0.400pt}}
\multiput(447.00,305.17)(3.000,-1.000){2}{\rule{0.723pt}{0.400pt}}
\put(453,303.67){\rule{1.445pt}{0.400pt}}
\multiput(453.00,304.17)(3.000,-1.000){2}{\rule{0.723pt}{0.400pt}}
\put(459,302.67){\rule{1.445pt}{0.400pt}}
\multiput(459.00,303.17)(3.000,-1.000){2}{\rule{0.723pt}{0.400pt}}
\put(465,301.17){\rule{1.300pt}{0.400pt}}
\multiput(465.00,302.17)(3.302,-2.000){2}{\rule{0.650pt}{0.400pt}}
\put(471,299.17){\rule{1.300pt}{0.400pt}}
\multiput(471.00,300.17)(3.302,-2.000){2}{\rule{0.650pt}{0.400pt}}
\put(477,297.17){\rule{1.500pt}{0.400pt}}
\multiput(477.00,298.17)(3.887,-2.000){2}{\rule{0.750pt}{0.400pt}}
\put(484,295.17){\rule{1.300pt}{0.400pt}}
\multiput(484.00,296.17)(3.302,-2.000){2}{\rule{0.650pt}{0.400pt}}
\multiput(490.00,293.95)(1.132,-0.447){3}{\rule{0.900pt}{0.108pt}}
\multiput(490.00,294.17)(4.132,-3.000){2}{\rule{0.450pt}{0.400pt}}
\multiput(496.00,290.95)(1.132,-0.447){3}{\rule{0.900pt}{0.108pt}}
\multiput(496.00,291.17)(4.132,-3.000){2}{\rule{0.450pt}{0.400pt}}
\multiput(502.00,287.94)(0.774,-0.468){5}{\rule{0.700pt}{0.113pt}}
\multiput(502.00,288.17)(4.547,-4.000){2}{\rule{0.350pt}{0.400pt}}
\multiput(508.00,283.93)(0.599,-0.477){7}{\rule{0.580pt}{0.115pt}}
\multiput(508.00,284.17)(4.796,-5.000){2}{\rule{0.290pt}{0.400pt}}
\multiput(514.00,278.93)(0.492,-0.485){11}{\rule{0.500pt}{0.117pt}}
\multiput(514.00,279.17)(5.962,-7.000){2}{\rule{0.250pt}{0.400pt}}
\multiput(521.59,267.88)(0.482,-1.485){9}{\rule{0.116pt}{1.233pt}}
\multiput(520.17,270.44)(6.000,-14.440){2}{\rule{0.400pt}{0.617pt}}
\put(434.0,306.0){\rule[-0.200pt]{3.132pt}{0.400pt}}
\put(323,256){\usebox{\plotpoint}}
\multiput(323.59,250.60)(0.482,-1.575){9}{\rule{0.116pt}{1.300pt}}
\multiput(322.17,253.30)(6.000,-15.302){2}{\rule{0.400pt}{0.650pt}}
\multiput(329.00,236.93)(0.492,-0.485){11}{\rule{0.500pt}{0.117pt}}
\multiput(329.00,237.17)(5.962,-7.000){2}{\rule{0.250pt}{0.400pt}}
\multiput(336.00,229.93)(0.599,-0.477){7}{\rule{0.580pt}{0.115pt}}
\multiput(336.00,230.17)(4.796,-5.000){2}{\rule{0.290pt}{0.400pt}}
\multiput(342.00,224.94)(0.774,-0.468){5}{\rule{0.700pt}{0.113pt}}
\multiput(342.00,225.17)(4.547,-4.000){2}{\rule{0.350pt}{0.400pt}}
\multiput(348.00,220.95)(1.132,-0.447){3}{\rule{0.900pt}{0.108pt}}
\multiput(348.00,221.17)(4.132,-3.000){2}{\rule{0.450pt}{0.400pt}}
\multiput(354.00,217.95)(1.132,-0.447){3}{\rule{0.900pt}{0.108pt}}
\multiput(354.00,218.17)(4.132,-3.000){2}{\rule{0.450pt}{0.400pt}}
\put(360,214.17){\rule{1.300pt}{0.400pt}}
\multiput(360.00,215.17)(3.302,-2.000){2}{\rule{0.650pt}{0.400pt}}
\put(366,212.17){\rule{1.500pt}{0.400pt}}
\multiput(366.00,213.17)(3.887,-2.000){2}{\rule{0.750pt}{0.400pt}}
\put(373,210.17){\rule{1.300pt}{0.400pt}}
\multiput(373.00,211.17)(3.302,-2.000){2}{\rule{0.650pt}{0.400pt}}
\put(379,208.17){\rule{1.300pt}{0.400pt}}
\multiput(379.00,209.17)(3.302,-2.000){2}{\rule{0.650pt}{0.400pt}}
\put(385,206.67){\rule{1.445pt}{0.400pt}}
\multiput(385.00,207.17)(3.000,-1.000){2}{\rule{0.723pt}{0.400pt}}
\put(391,205.67){\rule{1.445pt}{0.400pt}}
\multiput(391.00,206.17)(3.000,-1.000){2}{\rule{0.723pt}{0.400pt}}
\put(397,204.67){\rule{1.445pt}{0.400pt}}
\multiput(397.00,205.17)(3.000,-1.000){2}{\rule{0.723pt}{0.400pt}}
\put(416,203.67){\rule{1.445pt}{0.400pt}}
\multiput(416.00,204.17)(3.000,-1.000){2}{\rule{0.723pt}{0.400pt}}
\put(403.0,205.0){\rule[-0.200pt]{3.132pt}{0.400pt}}
\put(428,203.67){\rule{1.445pt}{0.400pt}}
\multiput(428.00,203.17)(3.000,1.000){2}{\rule{0.723pt}{0.400pt}}
\put(422.0,204.0){\rule[-0.200pt]{1.445pt}{0.400pt}}
\put(447,204.67){\rule{1.445pt}{0.400pt}}
\multiput(447.00,204.17)(3.000,1.000){2}{\rule{0.723pt}{0.400pt}}
\put(453,205.67){\rule{1.445pt}{0.400pt}}
\multiput(453.00,205.17)(3.000,1.000){2}{\rule{0.723pt}{0.400pt}}
\put(459,206.67){\rule{1.445pt}{0.400pt}}
\multiput(459.00,206.17)(3.000,1.000){2}{\rule{0.723pt}{0.400pt}}
\put(465,208.17){\rule{1.300pt}{0.400pt}}
\multiput(465.00,207.17)(3.302,2.000){2}{\rule{0.650pt}{0.400pt}}
\put(471,210.17){\rule{1.300pt}{0.400pt}}
\multiput(471.00,209.17)(3.302,2.000){2}{\rule{0.650pt}{0.400pt}}
\put(477,212.17){\rule{1.500pt}{0.400pt}}
\multiput(477.00,211.17)(3.887,2.000){2}{\rule{0.750pt}{0.400pt}}
\put(484,214.17){\rule{1.300pt}{0.400pt}}
\multiput(484.00,213.17)(3.302,2.000){2}{\rule{0.650pt}{0.400pt}}
\multiput(490.00,216.61)(1.132,0.447){3}{\rule{0.900pt}{0.108pt}}
\multiput(490.00,215.17)(4.132,3.000){2}{\rule{0.450pt}{0.400pt}}
\multiput(496.00,219.61)(1.132,0.447){3}{\rule{0.900pt}{0.108pt}}
\multiput(496.00,218.17)(4.132,3.000){2}{\rule{0.450pt}{0.400pt}}
\multiput(502.00,222.60)(0.774,0.468){5}{\rule{0.700pt}{0.113pt}}
\multiput(502.00,221.17)(4.547,4.000){2}{\rule{0.350pt}{0.400pt}}
\multiput(508.00,226.59)(0.599,0.477){7}{\rule{0.580pt}{0.115pt}}
\multiput(508.00,225.17)(4.796,5.000){2}{\rule{0.290pt}{0.400pt}}
\multiput(514.00,231.59)(0.492,0.485){11}{\rule{0.500pt}{0.117pt}}
\multiput(514.00,230.17)(5.962,7.000){2}{\rule{0.250pt}{0.400pt}}
\multiput(521.59,238.00)(0.482,1.575){9}{\rule{0.116pt}{1.300pt}}
\multiput(520.17,238.00)(6.000,15.302){2}{\rule{0.400pt}{0.650pt}}
\put(434.0,205.0){\rule[-0.200pt]{3.132pt}{0.400pt}}
\end{picture}
\vskip-1.8cm {\it Fig02: Diagrammatic representation for the equation (18).}
\end{center}

In the last step we have defined the two-point function 
\begin{equation}
T_{\mu \nu }^{AV}(k_{1},k_{2};m)=\int \frac{d^{4}k}{(2\pi )^{4}}Tr\left\{
i\gamma _{\mu }\gamma _{5}\frac{1}{[(\not{k}+{\not{k}}_{1})-m]}\gamma _{\nu }%
\frac{1}{[(\not{k}+{\not{k}}_{2})-m]}\right\} .
\end{equation}
Following this procedure, we also note the identity 
\begin{eqnarray}
&&(k_{3}-k_{1})_{\mu }\left\{ \gamma _{\nu }\frac{1}{(\not{k}+{\not{k}}%
_{2})-m}i\gamma _{\lambda }\gamma _{5}\frac{1}{(\not{k}+{\not{k}}_{3})-m}%
\gamma _{\mu }\frac{1}{(\not{k}+{\not{k}}_{1})-m}\right\} =  \nonumber \\
&&\;\;\;\;\;\;\;\;\;\;\;\;\left\{ \gamma _{\nu }\frac{1}{(\not{k}+{\not{k}}%
_{2})-m}i\gamma _{\lambda }\gamma _{5}\frac{1}{(\not{k}+{\not{k}}_{1})-m}%
\right\} -\left\{ \gamma _{\nu }\frac{1}{(\not{k}+{\not{k}}_{2})-m}i\gamma
_{\lambda }\gamma _{5}\frac{1}{(\not{k}+{\not{k}}_{3})-m}\right\} ,
\end{eqnarray}
and, consequently,(see fig03) 
\begin{equation}
(k_{3}-k_{1})_{\mu }T_{{\lambda }{\mu }{\nu }}^{AVV}=T_{{\lambda \nu }%
}^{AV}(k_{1},k_{2};m)-T_{{\lambda \nu }}^{AV}(k_{3},k_{2};m).
\end{equation}
Also, in a similar way, we get (see fig04) 
\begin{equation}
(k_{1}-k_{2})_{\nu }T_{{\lambda }{\mu }{\nu }}^{AVV}=T_{{\lambda \mu }%
}^{AV}(k_{3},k_{2};m)-T_{{\lambda \mu }}^{AV}(k_{3},k_{1};m).
\end{equation}

\begin{center}
\setlength{\unitlength}{0.240900pt} 
\begin{picture}(5,5)(-90,0)
\font\gnuplot=cmr10 at 10pt
\gnuplot
\put(480,235){\line(-1,0){10}}
\put(480,445){\line(-1,0){10}}
\put(480,235){\line(0,1){210}}
\put(80,235){\line(1,0){10}}
\put(80,445){\line(1,0){10}}
\put(80,235){\line(0,1){210}}
\put(70,345){\makebox(0,0)[r]{$(k_3-k_1)_\mu$}}
\put(90,345){\makebox(0,0)[l]{$i\gamma_\lambda \gamma_5$}}
\put(350,408){\makebox(0,0)[l]{$\gamma_\mu$}}
\put(350,272){\makebox(0,0)[l]{$\gamma_\nu$}}
\put(170,405){\makebox(0,0)[l]{$k+k_3$}}
\put(355,340){\makebox(0,0)[l]{$k+k_1$}}
\put(170,290){\makebox(0,0)[l]{$k+k_2$}}
\put(260,375){\vector(2,1){20}}
\put(280,305){\vector(-2,1){20}}
\put(341,350){\vector(0,-1){20}}
\put(325,400){$\bullet$}
\put(327,264){$\bullet$}
\put(188,333){$\bullet$}
\put(341,275){\line(0,1){140}}
\put(200,345){\line(2,1){140}}
\put(200,345){\line(2,-1){140}}
\put(510,325){$=$}
\end{picture}
\setlength{\unitlength}{0.240900pt} \ifx\plotpoint\undefined%
\newsavebox{\plotpoint}\fi
\sbox{\plotpoint}{\rule[-0.200pt]{0.400pt}{0.400pt}}%
\begin{picture}(750,450)(-400,-80)
\font\gnuplot=cmr10 at 10pt
\gnuplot
\put(313,256){\makebox(0,0)[l]{$\bullet$}}
\put(517,256){\makebox(0,0)[l]{$\bullet$}}
\put(425,204){\makebox(0,0)[l]{\vector(-1,0){20}}}
\put(425,327){\makebox(0,0)[l]{\vector(1,0){20}}}
\put(374,332){\makebox(0,0)[l]{$k+k_1$}}
\put(374,179){\makebox(0,0)[l]{$k+k_2$}}
\put(212,256){\makebox(0,0)[l]{$i\gamma_\lambda \gamma_5$}}
\put(547,256){\makebox(0,0)[l]{$\gamma_\nu$}}
\multiput(323.59,256.00)(0.482,1.485){9}{\rule{0.116pt}{1.233pt}}
\multiput(322.17,256.00)(6.000,14.440){2}{\rule{0.400pt}{0.617pt}}
\multiput(329.00,273.59)(0.492,0.485){11}{\rule{0.500pt}{0.117pt}}
\multiput(329.00,272.17)(5.962,7.000){2}{\rule{0.250pt}{0.400pt}}
\multiput(336.00,280.59)(0.599,0.477){7}{\rule{0.580pt}{0.115pt}}
\multiput(336.00,279.17)(4.796,5.000){2}{\rule{0.290pt}{0.400pt}}
\multiput(342.00,285.60)(0.774,0.468){5}{\rule{0.700pt}{0.113pt}}
\multiput(342.00,284.17)(4.547,4.000){2}{\rule{0.350pt}{0.400pt}}
\multiput(348.00,289.61)(1.132,0.447){3}{\rule{0.900pt}{0.108pt}}
\multiput(348.00,288.17)(4.132,3.000){2}{\rule{0.450pt}{0.400pt}}
\multiput(354.00,292.61)(1.132,0.447){3}{\rule{0.900pt}{0.108pt}}
\multiput(354.00,291.17)(4.132,3.000){2}{\rule{0.450pt}{0.400pt}}
\put(360,295.17){\rule{1.300pt}{0.400pt}}
\multiput(360.00,294.17)(3.302,2.000){2}{\rule{0.650pt}{0.400pt}}
\put(366,297.17){\rule{1.500pt}{0.400pt}}
\multiput(366.00,296.17)(3.887,2.000){2}{\rule{0.750pt}{0.400pt}}
\put(373,299.17){\rule{1.300pt}{0.400pt}}
\multiput(373.00,298.17)(3.302,2.000){2}{\rule{0.650pt}{0.400pt}}
\put(379,301.17){\rule{1.300pt}{0.400pt}}
\multiput(379.00,300.17)(3.302,2.000){2}{\rule{0.650pt}{0.400pt}}
\put(385,302.67){\rule{1.445pt}{0.400pt}}
\multiput(385.00,302.17)(3.000,1.000){2}{\rule{0.723pt}{0.400pt}}
\put(391,303.67){\rule{1.445pt}{0.400pt}}
\multiput(391.00,303.17)(3.000,1.000){2}{\rule{0.723pt}{0.400pt}}
\put(397,304.67){\rule{1.445pt}{0.400pt}}
\multiput(397.00,304.17)(3.000,1.000){2}{\rule{0.723pt}{0.400pt}}
\put(416,305.67){\rule{1.445pt}{0.400pt}}
\multiput(416.00,305.17)(3.000,1.000){2}{\rule{0.723pt}{0.400pt}}
\put(403.0,306.0){\rule[-0.200pt]{3.132pt}{0.400pt}}
\put(428,305.67){\rule{1.445pt}{0.400pt}}
\multiput(428.00,306.17)(3.000,-1.000){2}{\rule{0.723pt}{0.400pt}}
\put(422.0,307.0){\rule[-0.200pt]{1.445pt}{0.400pt}}
\put(447,304.67){\rule{1.445pt}{0.400pt}}
\multiput(447.00,305.17)(3.000,-1.000){2}{\rule{0.723pt}{0.400pt}}
\put(453,303.67){\rule{1.445pt}{0.400pt}}
\multiput(453.00,304.17)(3.000,-1.000){2}{\rule{0.723pt}{0.400pt}}
\put(459,302.67){\rule{1.445pt}{0.400pt}}
\multiput(459.00,303.17)(3.000,-1.000){2}{\rule{0.723pt}{0.400pt}}
\put(465,301.17){\rule{1.300pt}{0.400pt}}
\multiput(465.00,302.17)(3.302,-2.000){2}{\rule{0.650pt}{0.400pt}}
\put(471,299.17){\rule{1.300pt}{0.400pt}}
\multiput(471.00,300.17)(3.302,-2.000){2}{\rule{0.650pt}{0.400pt}}
\put(477,297.17){\rule{1.500pt}{0.400pt}}
\multiput(477.00,298.17)(3.887,-2.000){2}{\rule{0.750pt}{0.400pt}}
\put(484,295.17){\rule{1.300pt}{0.400pt}}
\multiput(484.00,296.17)(3.302,-2.000){2}{\rule{0.650pt}{0.400pt}}
\multiput(490.00,293.95)(1.132,-0.447){3}{\rule{0.900pt}{0.108pt}}
\multiput(490.00,294.17)(4.132,-3.000){2}{\rule{0.450pt}{0.400pt}}
\multiput(496.00,290.95)(1.132,-0.447){3}{\rule{0.900pt}{0.108pt}}
\multiput(496.00,291.17)(4.132,-3.000){2}{\rule{0.450pt}{0.400pt}}
\multiput(502.00,287.94)(0.774,-0.468){5}{\rule{0.700pt}{0.113pt}}
\multiput(502.00,288.17)(4.547,-4.000){2}{\rule{0.350pt}{0.400pt}}
\multiput(508.00,283.93)(0.599,-0.477){7}{\rule{0.580pt}{0.115pt}}
\multiput(508.00,284.17)(4.796,-5.000){2}{\rule{0.290pt}{0.400pt}}
\multiput(514.00,278.93)(0.492,-0.485){11}{\rule{0.500pt}{0.117pt}}
\multiput(514.00,279.17)(5.962,-7.000){2}{\rule{0.250pt}{0.400pt}}
\multiput(521.59,267.88)(0.482,-1.485){9}{\rule{0.116pt}{1.233pt}}
\multiput(520.17,270.44)(6.000,-14.440){2}{\rule{0.400pt}{0.617pt}}
\put(434.0,306.0){\rule[-0.200pt]{3.132pt}{0.400pt}}
\put(323,256){\usebox{\plotpoint}}
\multiput(323.59,250.60)(0.482,-1.575){9}{\rule{0.116pt}{1.300pt}}
\multiput(322.17,253.30)(6.000,-15.302){2}{\rule{0.400pt}{0.650pt}}
\multiput(329.00,236.93)(0.492,-0.485){11}{\rule{0.500pt}{0.117pt}}
\multiput(329.00,237.17)(5.962,-7.000){2}{\rule{0.250pt}{0.400pt}}
\multiput(336.00,229.93)(0.599,-0.477){7}{\rule{0.580pt}{0.115pt}}
\multiput(336.00,230.17)(4.796,-5.000){2}{\rule{0.290pt}{0.400pt}}
\multiput(342.00,224.94)(0.774,-0.468){5}{\rule{0.700pt}{0.113pt}}
\multiput(342.00,225.17)(4.547,-4.000){2}{\rule{0.350pt}{0.400pt}}
\multiput(348.00,220.95)(1.132,-0.447){3}{\rule{0.900pt}{0.108pt}}
\multiput(348.00,221.17)(4.132,-3.000){2}{\rule{0.450pt}{0.400pt}}
\multiput(354.00,217.95)(1.132,-0.447){3}{\rule{0.900pt}{0.108pt}}
\multiput(354.00,218.17)(4.132,-3.000){2}{\rule{0.450pt}{0.400pt}}
\put(360,214.17){\rule{1.300pt}{0.400pt}}
\multiput(360.00,215.17)(3.302,-2.000){2}{\rule{0.650pt}{0.400pt}}
\put(366,212.17){\rule{1.500pt}{0.400pt}}
\multiput(366.00,213.17)(3.887,-2.000){2}{\rule{0.750pt}{0.400pt}}
\put(373,210.17){\rule{1.300pt}{0.400pt}}
\multiput(373.00,211.17)(3.302,-2.000){2}{\rule{0.650pt}{0.400pt}}
\put(379,208.17){\rule{1.300pt}{0.400pt}}
\multiput(379.00,209.17)(3.302,-2.000){2}{\rule{0.650pt}{0.400pt}}
\put(385,206.67){\rule{1.445pt}{0.400pt}}
\multiput(385.00,207.17)(3.000,-1.000){2}{\rule{0.723pt}{0.400pt}}
\put(391,205.67){\rule{1.445pt}{0.400pt}}
\multiput(391.00,206.17)(3.000,-1.000){2}{\rule{0.723pt}{0.400pt}}
\put(397,204.67){\rule{1.445pt}{0.400pt}}
\multiput(397.00,205.17)(3.000,-1.000){2}{\rule{0.723pt}{0.400pt}}
\put(416,203.67){\rule{1.445pt}{0.400pt}}
\multiput(416.00,204.17)(3.000,-1.000){2}{\rule{0.723pt}{0.400pt}}
\put(403.0,205.0){\rule[-0.200pt]{3.132pt}{0.400pt}}
\put(428,203.67){\rule{1.445pt}{0.400pt}}
\multiput(428.00,203.17)(3.000,1.000){2}{\rule{0.723pt}{0.400pt}}
\put(422.0,204.0){\rule[-0.200pt]{1.445pt}{0.400pt}}
\put(447,204.67){\rule{1.445pt}{0.400pt}}
\multiput(447.00,204.17)(3.000,1.000){2}{\rule{0.723pt}{0.400pt}}
\put(453,205.67){\rule{1.445pt}{0.400pt}}
\multiput(453.00,205.17)(3.000,1.000){2}{\rule{0.723pt}{0.400pt}}
\put(459,206.67){\rule{1.445pt}{0.400pt}}
\multiput(459.00,206.17)(3.000,1.000){2}{\rule{0.723pt}{0.400pt}}
\put(465,208.17){\rule{1.300pt}{0.400pt}}
\multiput(465.00,207.17)(3.302,2.000){2}{\rule{0.650pt}{0.400pt}}
\put(471,210.17){\rule{1.300pt}{0.400pt}}
\multiput(471.00,209.17)(3.302,2.000){2}{\rule{0.650pt}{0.400pt}}
\put(477,212.17){\rule{1.500pt}{0.400pt}}
\multiput(477.00,211.17)(3.887,2.000){2}{\rule{0.750pt}{0.400pt}}
\put(484,214.17){\rule{1.300pt}{0.400pt}}
\multiput(484.00,213.17)(3.302,2.000){2}{\rule{0.650pt}{0.400pt}}
\multiput(490.00,216.61)(1.132,0.447){3}{\rule{0.900pt}{0.108pt}}
\multiput(490.00,215.17)(4.132,3.000){2}{\rule{0.450pt}{0.400pt}}
\multiput(496.00,219.61)(1.132,0.447){3}{\rule{0.900pt}{0.108pt}}
\multiput(496.00,218.17)(4.132,3.000){2}{\rule{0.450pt}{0.400pt}}
\multiput(502.00,222.60)(0.774,0.468){5}{\rule{0.700pt}{0.113pt}}
\multiput(502.00,221.17)(4.547,4.000){2}{\rule{0.350pt}{0.400pt}}
\multiput(508.00,226.59)(0.599,0.477){7}{\rule{0.580pt}{0.115pt}}
\multiput(508.00,225.17)(4.796,5.000){2}{\rule{0.290pt}{0.400pt}}
\multiput(514.00,231.59)(0.492,0.485){11}{\rule{0.500pt}{0.117pt}}
\multiput(514.00,230.17)(5.962,7.000){2}{\rule{0.250pt}{0.400pt}}
\multiput(521.59,238.00)(0.482,1.575){9}{\rule{0.116pt}{1.300pt}}
\multiput(520.17,238.00)(6.000,15.302){2}{\rule{0.400pt}{0.650pt}}
\put(434.0,205.0){\rule[-0.200pt]{3.132pt}{0.400pt}}
\put(607,255){\makebox(0,0)[l]{$-$}}
\end{picture}
\setlength{\unitlength}{0.240900pt} \ifx\plotpoint\undefined%
\newsavebox{\plotpoint}\fi
\sbox{\plotpoint}{\rule[-0.200pt]{0.400pt}{0.400pt}}%
\begin{picture}(750,450)(-80,-80)
\font\gnuplot=cmr10 at 10pt
\gnuplot
\put(313,256){\makebox(0,0)[l]{$\bullet$}}
\put(517,256){\makebox(0,0)[l]{$\bullet$}}
\put(425,204){\makebox(0,0)[l]{\vector(-1,0){20}}}
\put(425,327){\makebox(0,0)[l]{\vector(1,0){20}}}
\put(374,332){\makebox(0,0)[l]{$k+k_3$}}
\put(374,179){\makebox(0,0)[l]{$k+k_2$}}
\put(212,256){\makebox(0,0)[l]{$i\gamma_\lambda \gamma_5$}}
\put(547,256){\makebox(0,0)[l]{$\gamma_\nu$}}
\multiput(323.59,256.00)(0.482,1.485){9}{\rule{0.116pt}{1.233pt}}
\multiput(322.17,256.00)(6.000,14.440){2}{\rule{0.400pt}{0.617pt}}
\multiput(329.00,273.59)(0.492,0.485){11}{\rule{0.500pt}{0.117pt}}
\multiput(329.00,272.17)(5.962,7.000){2}{\rule{0.250pt}{0.400pt}}
\multiput(336.00,280.59)(0.599,0.477){7}{\rule{0.580pt}{0.115pt}}
\multiput(336.00,279.17)(4.796,5.000){2}{\rule{0.290pt}{0.400pt}}
\multiput(342.00,285.60)(0.774,0.468){5}{\rule{0.700pt}{0.113pt}}
\multiput(342.00,284.17)(4.547,4.000){2}{\rule{0.350pt}{0.400pt}}
\multiput(348.00,289.61)(1.132,0.447){3}{\rule{0.900pt}{0.108pt}}
\multiput(348.00,288.17)(4.132,3.000){2}{\rule{0.450pt}{0.400pt}}
\multiput(354.00,292.61)(1.132,0.447){3}{\rule{0.900pt}{0.108pt}}
\multiput(354.00,291.17)(4.132,3.000){2}{\rule{0.450pt}{0.400pt}}
\put(360,295.17){\rule{1.300pt}{0.400pt}}
\multiput(360.00,294.17)(3.302,2.000){2}{\rule{0.650pt}{0.400pt}}
\put(366,297.17){\rule{1.500pt}{0.400pt}}
\multiput(366.00,296.17)(3.887,2.000){2}{\rule{0.750pt}{0.400pt}}
\put(373,299.17){\rule{1.300pt}{0.400pt}}
\multiput(373.00,298.17)(3.302,2.000){2}{\rule{0.650pt}{0.400pt}}
\put(379,301.17){\rule{1.300pt}{0.400pt}}
\multiput(379.00,300.17)(3.302,2.000){2}{\rule{0.650pt}{0.400pt}}
\put(385,302.67){\rule{1.445pt}{0.400pt}}
\multiput(385.00,302.17)(3.000,1.000){2}{\rule{0.723pt}{0.400pt}}
\put(391,303.67){\rule{1.445pt}{0.400pt}}
\multiput(391.00,303.17)(3.000,1.000){2}{\rule{0.723pt}{0.400pt}}
\put(397,304.67){\rule{1.445pt}{0.400pt}}
\multiput(397.00,304.17)(3.000,1.000){2}{\rule{0.723pt}{0.400pt}}
\put(416,305.67){\rule{1.445pt}{0.400pt}}
\multiput(416.00,305.17)(3.000,1.000){2}{\rule{0.723pt}{0.400pt}}
\put(403.0,306.0){\rule[-0.200pt]{3.132pt}{0.400pt}}
\put(428,305.67){\rule{1.445pt}{0.400pt}}
\multiput(428.00,306.17)(3.000,-1.000){2}{\rule{0.723pt}{0.400pt}}
\put(422.0,307.0){\rule[-0.200pt]{1.445pt}{0.400pt}}
\put(447,304.67){\rule{1.445pt}{0.400pt}}
\multiput(447.00,305.17)(3.000,-1.000){2}{\rule{0.723pt}{0.400pt}}
\put(453,303.67){\rule{1.445pt}{0.400pt}}
\multiput(453.00,304.17)(3.000,-1.000){2}{\rule{0.723pt}{0.400pt}}
\put(459,302.67){\rule{1.445pt}{0.400pt}}
\multiput(459.00,303.17)(3.000,-1.000){2}{\rule{0.723pt}{0.400pt}}
\put(465,301.17){\rule{1.300pt}{0.400pt}}
\multiput(465.00,302.17)(3.302,-2.000){2}{\rule{0.650pt}{0.400pt}}
\put(471,299.17){\rule{1.300pt}{0.400pt}}
\multiput(471.00,300.17)(3.302,-2.000){2}{\rule{0.650pt}{0.400pt}}
\put(477,297.17){\rule{1.500pt}{0.400pt}}
\multiput(477.00,298.17)(3.887,-2.000){2}{\rule{0.750pt}{0.400pt}}
\put(484,295.17){\rule{1.300pt}{0.400pt}}
\multiput(484.00,296.17)(3.302,-2.000){2}{\rule{0.650pt}{0.400pt}}
\multiput(490.00,293.95)(1.132,-0.447){3}{\rule{0.900pt}{0.108pt}}
\multiput(490.00,294.17)(4.132,-3.000){2}{\rule{0.450pt}{0.400pt}}
\multiput(496.00,290.95)(1.132,-0.447){3}{\rule{0.900pt}{0.108pt}}
\multiput(496.00,291.17)(4.132,-3.000){2}{\rule{0.450pt}{0.400pt}}
\multiput(502.00,287.94)(0.774,-0.468){5}{\rule{0.700pt}{0.113pt}}
\multiput(502.00,288.17)(4.547,-4.000){2}{\rule{0.350pt}{0.400pt}}
\multiput(508.00,283.93)(0.599,-0.477){7}{\rule{0.580pt}{0.115pt}}
\multiput(508.00,284.17)(4.796,-5.000){2}{\rule{0.290pt}{0.400pt}}
\multiput(514.00,278.93)(0.492,-0.485){11}{\rule{0.500pt}{0.117pt}}
\multiput(514.00,279.17)(5.962,-7.000){2}{\rule{0.250pt}{0.400pt}}
\multiput(521.59,267.88)(0.482,-1.485){9}{\rule{0.116pt}{1.233pt}}
\multiput(520.17,270.44)(6.000,-14.440){2}{\rule{0.400pt}{0.617pt}}
\put(434.0,306.0){\rule[-0.200pt]{3.132pt}{0.400pt}}
\put(323,256){\usebox{\plotpoint}}
\multiput(323.59,250.60)(0.482,-1.575){9}{\rule{0.116pt}{1.300pt}}
\multiput(322.17,253.30)(6.000,-15.302){2}{\rule{0.400pt}{0.650pt}}
\multiput(329.00,236.93)(0.492,-0.485){11}{\rule{0.500pt}{0.117pt}}
\multiput(329.00,237.17)(5.962,-7.000){2}{\rule{0.250pt}{0.400pt}}
\multiput(336.00,229.93)(0.599,-0.477){7}{\rule{0.580pt}{0.115pt}}
\multiput(336.00,230.17)(4.796,-5.000){2}{\rule{0.290pt}{0.400pt}}
\multiput(342.00,224.94)(0.774,-0.468){5}{\rule{0.700pt}{0.113pt}}
\multiput(342.00,225.17)(4.547,-4.000){2}{\rule{0.350pt}{0.400pt}}
\multiput(348.00,220.95)(1.132,-0.447){3}{\rule{0.900pt}{0.108pt}}
\multiput(348.00,221.17)(4.132,-3.000){2}{\rule{0.450pt}{0.400pt}}
\multiput(354.00,217.95)(1.132,-0.447){3}{\rule{0.900pt}{0.108pt}}
\multiput(354.00,218.17)(4.132,-3.000){2}{\rule{0.450pt}{0.400pt}}
\put(360,214.17){\rule{1.300pt}{0.400pt}}
\multiput(360.00,215.17)(3.302,-2.000){2}{\rule{0.650pt}{0.400pt}}
\put(366,212.17){\rule{1.500pt}{0.400pt}}
\multiput(366.00,213.17)(3.887,-2.000){2}{\rule{0.750pt}{0.400pt}}
\put(373,210.17){\rule{1.300pt}{0.400pt}}
\multiput(373.00,211.17)(3.302,-2.000){2}{\rule{0.650pt}{0.400pt}}
\put(379,208.17){\rule{1.300pt}{0.400pt}}
\multiput(379.00,209.17)(3.302,-2.000){2}{\rule{0.650pt}{0.400pt}}
\put(385,206.67){\rule{1.445pt}{0.400pt}}
\multiput(385.00,207.17)(3.000,-1.000){2}{\rule{0.723pt}{0.400pt}}
\put(391,205.67){\rule{1.445pt}{0.400pt}}
\multiput(391.00,206.17)(3.000,-1.000){2}{\rule{0.723pt}{0.400pt}}
\put(397,204.67){\rule{1.445pt}{0.400pt}}
\multiput(397.00,205.17)(3.000,-1.000){2}{\rule{0.723pt}{0.400pt}}
\put(416,203.67){\rule{1.445pt}{0.400pt}}
\multiput(416.00,204.17)(3.000,-1.000){2}{\rule{0.723pt}{0.400pt}}
\put(403.0,205.0){\rule[-0.200pt]{3.132pt}{0.400pt}}
\put(428,203.67){\rule{1.445pt}{0.400pt}}
\multiput(428.00,203.17)(3.000,1.000){2}{\rule{0.723pt}{0.400pt}}
\put(422.0,204.0){\rule[-0.200pt]{1.445pt}{0.400pt}}
\put(447,204.67){\rule{1.445pt}{0.400pt}}
\multiput(447.00,204.17)(3.000,1.000){2}{\rule{0.723pt}{0.400pt}}
\put(453,205.67){\rule{1.445pt}{0.400pt}}
\multiput(453.00,205.17)(3.000,1.000){2}{\rule{0.723pt}{0.400pt}}
\put(459,206.67){\rule{1.445pt}{0.400pt}}
\multiput(459.00,206.17)(3.000,1.000){2}{\rule{0.723pt}{0.400pt}}
\put(465,208.17){\rule{1.300pt}{0.400pt}}
\multiput(465.00,207.17)(3.302,2.000){2}{\rule{0.650pt}{0.400pt}}
\put(471,210.17){\rule{1.300pt}{0.400pt}}
\multiput(471.00,209.17)(3.302,2.000){2}{\rule{0.650pt}{0.400pt}}
\put(477,212.17){\rule{1.500pt}{0.400pt}}
\multiput(477.00,211.17)(3.887,2.000){2}{\rule{0.750pt}{0.400pt}}
\put(484,214.17){\rule{1.300pt}{0.400pt}}
\multiput(484.00,213.17)(3.302,2.000){2}{\rule{0.650pt}{0.400pt}}
\multiput(490.00,216.61)(1.132,0.447){3}{\rule{0.900pt}{0.108pt}}
\multiput(490.00,215.17)(4.132,3.000){2}{\rule{0.450pt}{0.400pt}}
\multiput(496.00,219.61)(1.132,0.447){3}{\rule{0.900pt}{0.108pt}}
\multiput(496.00,218.17)(4.132,3.000){2}{\rule{0.450pt}{0.400pt}}
\multiput(502.00,222.60)(0.774,0.468){5}{\rule{0.700pt}{0.113pt}}
\multiput(502.00,221.17)(4.547,4.000){2}{\rule{0.350pt}{0.400pt}}
\multiput(508.00,226.59)(0.599,0.477){7}{\rule{0.580pt}{0.115pt}}
\multiput(508.00,225.17)(4.796,5.000){2}{\rule{0.290pt}{0.400pt}}
\multiput(514.00,231.59)(0.492,0.485){11}{\rule{0.500pt}{0.117pt}}
\multiput(514.00,230.17)(5.962,7.000){2}{\rule{0.250pt}{0.400pt}}
\multiput(521.59,238.00)(0.482,1.575){9}{\rule{0.116pt}{1.300pt}}
\multiput(520.17,238.00)(6.000,15.302){2}{\rule{0.400pt}{0.650pt}}
\put(434.0,205.0){\rule[-0.200pt]{3.132pt}{0.400pt}}
\end{picture}
\vskip-1.8cm {\it Fig03: Diagrammatic representation for the identity (21).}
\end{center}


\begin{center}
\setlength{\unitlength}{0.240900pt} 
\begin{picture}(5,5)(-90,0)
\font\gnuplot=cmr10 at 10pt
\gnuplot
\put(480,235){\line(-1,0){10}}
\put(480,445){\line(-1,0){10}}
\put(480,235){\line(0,1){210}}
\put(80,235){\line(1,0){10}}
\put(80,445){\line(1,0){10}}
\put(80,235){\line(0,1){210}}
\put(70,345){\makebox(0,0)[r]{$(k_1-k_2)_\nu$}}
\put(90,345){\makebox(0,0)[l]{$i\gamma_\lambda \gamma_5$}}
\put(350,408){\makebox(0,0)[l]{$\gamma_\mu$}}
\put(350,272){\makebox(0,0)[l]{$\gamma_\nu$}}
\put(170,405){\makebox(0,0)[l]{$k+k_3$}}
\put(355,340){\makebox(0,0)[l]{$k+k_1$}}
\put(170,290){\makebox(0,0)[l]{$k+k_2$}}
\put(260,375){\vector(2,1){20}}
\put(280,305){\vector(-2,1){20}}
\put(341,350){\vector(0,-1){20}}
\put(325,400){$\bullet$}
\put(327,264){$\bullet$}
\put(188,333){$\bullet$}
\put(341,275){\line(0,1){140}}
\put(200,345){\line(2,1){140}}
\put(200,345){\line(2,-1){140}}
\put(510,325){$=$}
\end{picture}
\setlength{\unitlength}{0.240900pt} \ifx\plotpoint\undefined%
\newsavebox{\plotpoint}\fi
\sbox{\plotpoint}{\rule[-0.200pt]{0.400pt}{0.400pt}}%
\begin{picture}(750,450)(-400,-80)
\font\gnuplot=cmr10 at 10pt
\gnuplot
\put(313,256){\makebox(0,0)[l]{$\bullet$}}
\put(517,256){\makebox(0,0)[l]{$\bullet$}}
\put(425,204){\makebox(0,0)[l]{\vector(-1,0){20}}}
\put(425,327){\makebox(0,0)[l]{\vector(1,0){20}}}
\put(374,332){\makebox(0,0)[l]{$k+k_3$}}
\put(374,179){\makebox(0,0)[l]{$k+k_2$}}
\put(212,256){\makebox(0,0)[l]{$i\gamma_\lambda \gamma_5$}}
\put(547,256){\makebox(0,0)[l]{$\gamma_\mu$}}
\multiput(323.59,256.00)(0.482,1.485){9}{\rule{0.116pt}{1.233pt}}
\multiput(322.17,256.00)(6.000,14.440){2}{\rule{0.400pt}{0.617pt}}
\multiput(329.00,273.59)(0.492,0.485){11}{\rule{0.500pt}{0.117pt}}
\multiput(329.00,272.17)(5.962,7.000){2}{\rule{0.250pt}{0.400pt}}
\multiput(336.00,280.59)(0.599,0.477){7}{\rule{0.580pt}{0.115pt}}
\multiput(336.00,279.17)(4.796,5.000){2}{\rule{0.290pt}{0.400pt}}
\multiput(342.00,285.60)(0.774,0.468){5}{\rule{0.700pt}{0.113pt}}
\multiput(342.00,284.17)(4.547,4.000){2}{\rule{0.350pt}{0.400pt}}
\multiput(348.00,289.61)(1.132,0.447){3}{\rule{0.900pt}{0.108pt}}
\multiput(348.00,288.17)(4.132,3.000){2}{\rule{0.450pt}{0.400pt}}
\multiput(354.00,292.61)(1.132,0.447){3}{\rule{0.900pt}{0.108pt}}
\multiput(354.00,291.17)(4.132,3.000){2}{\rule{0.450pt}{0.400pt}}
\put(360,295.17){\rule{1.300pt}{0.400pt}}
\multiput(360.00,294.17)(3.302,2.000){2}{\rule{0.650pt}{0.400pt}}
\put(366,297.17){\rule{1.500pt}{0.400pt}}
\multiput(366.00,296.17)(3.887,2.000){2}{\rule{0.750pt}{0.400pt}}
\put(373,299.17){\rule{1.300pt}{0.400pt}}
\multiput(373.00,298.17)(3.302,2.000){2}{\rule{0.650pt}{0.400pt}}
\put(379,301.17){\rule{1.300pt}{0.400pt}}
\multiput(379.00,300.17)(3.302,2.000){2}{\rule{0.650pt}{0.400pt}}
\put(385,302.67){\rule{1.445pt}{0.400pt}}
\multiput(385.00,302.17)(3.000,1.000){2}{\rule{0.723pt}{0.400pt}}
\put(391,303.67){\rule{1.445pt}{0.400pt}}
\multiput(391.00,303.17)(3.000,1.000){2}{\rule{0.723pt}{0.400pt}}
\put(397,304.67){\rule{1.445pt}{0.400pt}}
\multiput(397.00,304.17)(3.000,1.000){2}{\rule{0.723pt}{0.400pt}}
\put(416,305.67){\rule{1.445pt}{0.400pt}}
\multiput(416.00,305.17)(3.000,1.000){2}{\rule{0.723pt}{0.400pt}}
\put(403.0,306.0){\rule[-0.200pt]{3.132pt}{0.400pt}}
\put(428,305.67){\rule{1.445pt}{0.400pt}}
\multiput(428.00,306.17)(3.000,-1.000){2}{\rule{0.723pt}{0.400pt}}
\put(422.0,307.0){\rule[-0.200pt]{1.445pt}{0.400pt}}
\put(447,304.67){\rule{1.445pt}{0.400pt}}
\multiput(447.00,305.17)(3.000,-1.000){2}{\rule{0.723pt}{0.400pt}}
\put(453,303.67){\rule{1.445pt}{0.400pt}}
\multiput(453.00,304.17)(3.000,-1.000){2}{\rule{0.723pt}{0.400pt}}
\put(459,302.67){\rule{1.445pt}{0.400pt}}
\multiput(459.00,303.17)(3.000,-1.000){2}{\rule{0.723pt}{0.400pt}}
\put(465,301.17){\rule{1.300pt}{0.400pt}}
\multiput(465.00,302.17)(3.302,-2.000){2}{\rule{0.650pt}{0.400pt}}
\put(471,299.17){\rule{1.300pt}{0.400pt}}
\multiput(471.00,300.17)(3.302,-2.000){2}{\rule{0.650pt}{0.400pt}}
\put(477,297.17){\rule{1.500pt}{0.400pt}}
\multiput(477.00,298.17)(3.887,-2.000){2}{\rule{0.750pt}{0.400pt}}
\put(484,295.17){\rule{1.300pt}{0.400pt}}
\multiput(484.00,296.17)(3.302,-2.000){2}{\rule{0.650pt}{0.400pt}}
\multiput(490.00,293.95)(1.132,-0.447){3}{\rule{0.900pt}{0.108pt}}
\multiput(490.00,294.17)(4.132,-3.000){2}{\rule{0.450pt}{0.400pt}}
\multiput(496.00,290.95)(1.132,-0.447){3}{\rule{0.900pt}{0.108pt}}
\multiput(496.00,291.17)(4.132,-3.000){2}{\rule{0.450pt}{0.400pt}}
\multiput(502.00,287.94)(0.774,-0.468){5}{\rule{0.700pt}{0.113pt}}
\multiput(502.00,288.17)(4.547,-4.000){2}{\rule{0.350pt}{0.400pt}}
\multiput(508.00,283.93)(0.599,-0.477){7}{\rule{0.580pt}{0.115pt}}
\multiput(508.00,284.17)(4.796,-5.000){2}{\rule{0.290pt}{0.400pt}}
\multiput(514.00,278.93)(0.492,-0.485){11}{\rule{0.500pt}{0.117pt}}
\multiput(514.00,279.17)(5.962,-7.000){2}{\rule{0.250pt}{0.400pt}}
\multiput(521.59,267.88)(0.482,-1.485){9}{\rule{0.116pt}{1.233pt}}
\multiput(520.17,270.44)(6.000,-14.440){2}{\rule{0.400pt}{0.617pt}}
\put(434.0,306.0){\rule[-0.200pt]{3.132pt}{0.400pt}}
\put(323,256){\usebox{\plotpoint}}
\multiput(323.59,250.60)(0.482,-1.575){9}{\rule{0.116pt}{1.300pt}}
\multiput(322.17,253.30)(6.000,-15.302){2}{\rule{0.400pt}{0.650pt}}
\multiput(329.00,236.93)(0.492,-0.485){11}{\rule{0.500pt}{0.117pt}}
\multiput(329.00,237.17)(5.962,-7.000){2}{\rule{0.250pt}{0.400pt}}
\multiput(336.00,229.93)(0.599,-0.477){7}{\rule{0.580pt}{0.115pt}}
\multiput(336.00,230.17)(4.796,-5.000){2}{\rule{0.290pt}{0.400pt}}
\multiput(342.00,224.94)(0.774,-0.468){5}{\rule{0.700pt}{0.113pt}}
\multiput(342.00,225.17)(4.547,-4.000){2}{\rule{0.350pt}{0.400pt}}
\multiput(348.00,220.95)(1.132,-0.447){3}{\rule{0.900pt}{0.108pt}}
\multiput(348.00,221.17)(4.132,-3.000){2}{\rule{0.450pt}{0.400pt}}
\multiput(354.00,217.95)(1.132,-0.447){3}{\rule{0.900pt}{0.108pt}}
\multiput(354.00,218.17)(4.132,-3.000){2}{\rule{0.450pt}{0.400pt}}
\put(360,214.17){\rule{1.300pt}{0.400pt}}
\multiput(360.00,215.17)(3.302,-2.000){2}{\rule{0.650pt}{0.400pt}}
\put(366,212.17){\rule{1.500pt}{0.400pt}}
\multiput(366.00,213.17)(3.887,-2.000){2}{\rule{0.750pt}{0.400pt}}
\put(373,210.17){\rule{1.300pt}{0.400pt}}
\multiput(373.00,211.17)(3.302,-2.000){2}{\rule{0.650pt}{0.400pt}}
\put(379,208.17){\rule{1.300pt}{0.400pt}}
\multiput(379.00,209.17)(3.302,-2.000){2}{\rule{0.650pt}{0.400pt}}
\put(385,206.67){\rule{1.445pt}{0.400pt}}
\multiput(385.00,207.17)(3.000,-1.000){2}{\rule{0.723pt}{0.400pt}}
\put(391,205.67){\rule{1.445pt}{0.400pt}}
\multiput(391.00,206.17)(3.000,-1.000){2}{\rule{0.723pt}{0.400pt}}
\put(397,204.67){\rule{1.445pt}{0.400pt}}
\multiput(397.00,205.17)(3.000,-1.000){2}{\rule{0.723pt}{0.400pt}}
\put(416,203.67){\rule{1.445pt}{0.400pt}}
\multiput(416.00,204.17)(3.000,-1.000){2}{\rule{0.723pt}{0.400pt}}
\put(403.0,205.0){\rule[-0.200pt]{3.132pt}{0.400pt}}
\put(428,203.67){\rule{1.445pt}{0.400pt}}
\multiput(428.00,203.17)(3.000,1.000){2}{\rule{0.723pt}{0.400pt}}
\put(422.0,204.0){\rule[-0.200pt]{1.445pt}{0.400pt}}
\put(447,204.67){\rule{1.445pt}{0.400pt}}
\multiput(447.00,204.17)(3.000,1.000){2}{\rule{0.723pt}{0.400pt}}
\put(453,205.67){\rule{1.445pt}{0.400pt}}
\multiput(453.00,205.17)(3.000,1.000){2}{\rule{0.723pt}{0.400pt}}
\put(459,206.67){\rule{1.445pt}{0.400pt}}
\multiput(459.00,206.17)(3.000,1.000){2}{\rule{0.723pt}{0.400pt}}
\put(465,208.17){\rule{1.300pt}{0.400pt}}
\multiput(465.00,207.17)(3.302,2.000){2}{\rule{0.650pt}{0.400pt}}
\put(471,210.17){\rule{1.300pt}{0.400pt}}
\multiput(471.00,209.17)(3.302,2.000){2}{\rule{0.650pt}{0.400pt}}
\put(477,212.17){\rule{1.500pt}{0.400pt}}
\multiput(477.00,211.17)(3.887,2.000){2}{\rule{0.750pt}{0.400pt}}
\put(484,214.17){\rule{1.300pt}{0.400pt}}
\multiput(484.00,213.17)(3.302,2.000){2}{\rule{0.650pt}{0.400pt}}
\multiput(490.00,216.61)(1.132,0.447){3}{\rule{0.900pt}{0.108pt}}
\multiput(490.00,215.17)(4.132,3.000){2}{\rule{0.450pt}{0.400pt}}
\multiput(496.00,219.61)(1.132,0.447){3}{\rule{0.900pt}{0.108pt}}
\multiput(496.00,218.17)(4.132,3.000){2}{\rule{0.450pt}{0.400pt}}
\multiput(502.00,222.60)(0.774,0.468){5}{\rule{0.700pt}{0.113pt}}
\multiput(502.00,221.17)(4.547,4.000){2}{\rule{0.350pt}{0.400pt}}
\multiput(508.00,226.59)(0.599,0.477){7}{\rule{0.580pt}{0.115pt}}
\multiput(508.00,225.17)(4.796,5.000){2}{\rule{0.290pt}{0.400pt}}
\multiput(514.00,231.59)(0.492,0.485){11}{\rule{0.500pt}{0.117pt}}
\multiput(514.00,230.17)(5.962,7.000){2}{\rule{0.250pt}{0.400pt}}
\multiput(521.59,238.00)(0.482,1.575){9}{\rule{0.116pt}{1.300pt}}
\multiput(520.17,238.00)(6.000,15.302){2}{\rule{0.400pt}{0.650pt}}
\put(434.0,205.0){\rule[-0.200pt]{3.132pt}{0.400pt}}
\put(607,255){\makebox(0,0)[l]{$-$}}
\end{picture}
\setlength{\unitlength}{0.240900pt} \ifx\plotpoint\undefined%
\newsavebox{\plotpoint}\fi
\sbox{\plotpoint}{\rule[-0.200pt]{0.400pt}{0.400pt}}%
\begin{picture}(750,450)(-90,-80)
\font\gnuplot=cmr10 at 10pt
\gnuplot
\put(313,256){\makebox(0,0)[l]{$\bullet$}}
\put(517,256){\makebox(0,0)[l]{$\bullet$}}
\put(425,204){\makebox(0,0)[l]{\vector(-1,0){20}}}
\put(425,327){\makebox(0,0)[l]{\vector(1,0){20}}}
\put(374,332){\makebox(0,0)[l]{$k+k_3$}}
\put(374,179){\makebox(0,0)[l]{$k+k_1$}}
\put(212,256){\makebox(0,0)[l]{$i\gamma_\lambda \gamma_5$}}
\put(547,256){\makebox(0,0)[l]{$\gamma_\mu$}}
\multiput(323.59,256.00)(0.482,1.485){9}{\rule{0.116pt}{1.233pt}}
\multiput(322.17,256.00)(6.000,14.440){2}{\rule{0.400pt}{0.617pt}}
\multiput(329.00,273.59)(0.492,0.485){11}{\rule{0.500pt}{0.117pt}}
\multiput(329.00,272.17)(5.962,7.000){2}{\rule{0.250pt}{0.400pt}}
\multiput(336.00,280.59)(0.599,0.477){7}{\rule{0.580pt}{0.115pt}}
\multiput(336.00,279.17)(4.796,5.000){2}{\rule{0.290pt}{0.400pt}}
\multiput(342.00,285.60)(0.774,0.468){5}{\rule{0.700pt}{0.113pt}}
\multiput(342.00,284.17)(4.547,4.000){2}{\rule{0.350pt}{0.400pt}}
\multiput(348.00,289.61)(1.132,0.447){3}{\rule{0.900pt}{0.108pt}}
\multiput(348.00,288.17)(4.132,3.000){2}{\rule{0.450pt}{0.400pt}}
\multiput(354.00,292.61)(1.132,0.447){3}{\rule{0.900pt}{0.108pt}}
\multiput(354.00,291.17)(4.132,3.000){2}{\rule{0.450pt}{0.400pt}}
\put(360,295.17){\rule{1.300pt}{0.400pt}}
\multiput(360.00,294.17)(3.302,2.000){2}{\rule{0.650pt}{0.400pt}}
\put(366,297.17){\rule{1.500pt}{0.400pt}}
\multiput(366.00,296.17)(3.887,2.000){2}{\rule{0.750pt}{0.400pt}}
\put(373,299.17){\rule{1.300pt}{0.400pt}}
\multiput(373.00,298.17)(3.302,2.000){2}{\rule{0.650pt}{0.400pt}}
\put(379,301.17){\rule{1.300pt}{0.400pt}}
\multiput(379.00,300.17)(3.302,2.000){2}{\rule{0.650pt}{0.400pt}}
\put(385,302.67){\rule{1.445pt}{0.400pt}}
\multiput(385.00,302.17)(3.000,1.000){2}{\rule{0.723pt}{0.400pt}}
\put(391,303.67){\rule{1.445pt}{0.400pt}}
\multiput(391.00,303.17)(3.000,1.000){2}{\rule{0.723pt}{0.400pt}}
\put(397,304.67){\rule{1.445pt}{0.400pt}}
\multiput(397.00,304.17)(3.000,1.000){2}{\rule{0.723pt}{0.400pt}}
\put(416,305.67){\rule{1.445pt}{0.400pt}}
\multiput(416.00,305.17)(3.000,1.000){2}{\rule{0.723pt}{0.400pt}}
\put(403.0,306.0){\rule[-0.200pt]{3.132pt}{0.400pt}}
\put(428,305.67){\rule{1.445pt}{0.400pt}}
\multiput(428.00,306.17)(3.000,-1.000){2}{\rule{0.723pt}{0.400pt}}
\put(422.0,307.0){\rule[-0.200pt]{1.445pt}{0.400pt}}
\put(447,304.67){\rule{1.445pt}{0.400pt}}
\multiput(447.00,305.17)(3.000,-1.000){2}{\rule{0.723pt}{0.400pt}}
\put(453,303.67){\rule{1.445pt}{0.400pt}}
\multiput(453.00,304.17)(3.000,-1.000){2}{\rule{0.723pt}{0.400pt}}
\put(459,302.67){\rule{1.445pt}{0.400pt}}
\multiput(459.00,303.17)(3.000,-1.000){2}{\rule{0.723pt}{0.400pt}}
\put(465,301.17){\rule{1.300pt}{0.400pt}}
\multiput(465.00,302.17)(3.302,-2.000){2}{\rule{0.650pt}{0.400pt}}
\put(471,299.17){\rule{1.300pt}{0.400pt}}
\multiput(471.00,300.17)(3.302,-2.000){2}{\rule{0.650pt}{0.400pt}}
\put(477,297.17){\rule{1.500pt}{0.400pt}}
\multiput(477.00,298.17)(3.887,-2.000){2}{\rule{0.750pt}{0.400pt}}
\put(484,295.17){\rule{1.300pt}{0.400pt}}
\multiput(484.00,296.17)(3.302,-2.000){2}{\rule{0.650pt}{0.400pt}}
\multiput(490.00,293.95)(1.132,-0.447){3}{\rule{0.900pt}{0.108pt}}
\multiput(490.00,294.17)(4.132,-3.000){2}{\rule{0.450pt}{0.400pt}}
\multiput(496.00,290.95)(1.132,-0.447){3}{\rule{0.900pt}{0.108pt}}
\multiput(496.00,291.17)(4.132,-3.000){2}{\rule{0.450pt}{0.400pt}}
\multiput(502.00,287.94)(0.774,-0.468){5}{\rule{0.700pt}{0.113pt}}
\multiput(502.00,288.17)(4.547,-4.000){2}{\rule{0.350pt}{0.400pt}}
\multiput(508.00,283.93)(0.599,-0.477){7}{\rule{0.580pt}{0.115pt}}
\multiput(508.00,284.17)(4.796,-5.000){2}{\rule{0.290pt}{0.400pt}}
\multiput(514.00,278.93)(0.492,-0.485){11}{\rule{0.500pt}{0.117pt}}
\multiput(514.00,279.17)(5.962,-7.000){2}{\rule{0.250pt}{0.400pt}}
\multiput(521.59,267.88)(0.482,-1.485){9}{\rule{0.116pt}{1.233pt}}
\multiput(520.17,270.44)(6.000,-14.440){2}{\rule{0.400pt}{0.617pt}}
\put(434.0,306.0){\rule[-0.200pt]{3.132pt}{0.400pt}}
\put(323,256){\usebox{\plotpoint}}
\multiput(323.59,250.60)(0.482,-1.575){9}{\rule{0.116pt}{1.300pt}}
\multiput(322.17,253.30)(6.000,-15.302){2}{\rule{0.400pt}{0.650pt}}
\multiput(329.00,236.93)(0.492,-0.485){11}{\rule{0.500pt}{0.117pt}}
\multiput(329.00,237.17)(5.962,-7.000){2}{\rule{0.250pt}{0.400pt}}
\multiput(336.00,229.93)(0.599,-0.477){7}{\rule{0.580pt}{0.115pt}}
\multiput(336.00,230.17)(4.796,-5.000){2}{\rule{0.290pt}{0.400pt}}
\multiput(342.00,224.94)(0.774,-0.468){5}{\rule{0.700pt}{0.113pt}}
\multiput(342.00,225.17)(4.547,-4.000){2}{\rule{0.350pt}{0.400pt}}
\multiput(348.00,220.95)(1.132,-0.447){3}{\rule{0.900pt}{0.108pt}}
\multiput(348.00,221.17)(4.132,-3.000){2}{\rule{0.450pt}{0.400pt}}
\multiput(354.00,217.95)(1.132,-0.447){3}{\rule{0.900pt}{0.108pt}}
\multiput(354.00,218.17)(4.132,-3.000){2}{\rule{0.450pt}{0.400pt}}
\put(360,214.17){\rule{1.300pt}{0.400pt}}
\multiput(360.00,215.17)(3.302,-2.000){2}{\rule{0.650pt}{0.400pt}}
\put(366,212.17){\rule{1.500pt}{0.400pt}}
\multiput(366.00,213.17)(3.887,-2.000){2}{\rule{0.750pt}{0.400pt}}
\put(373,210.17){\rule{1.300pt}{0.400pt}}
\multiput(373.00,211.17)(3.302,-2.000){2}{\rule{0.650pt}{0.400pt}}
\put(379,208.17){\rule{1.300pt}{0.400pt}}
\multiput(379.00,209.17)(3.302,-2.000){2}{\rule{0.650pt}{0.400pt}}
\put(385,206.67){\rule{1.445pt}{0.400pt}}
\multiput(385.00,207.17)(3.000,-1.000){2}{\rule{0.723pt}{0.400pt}}
\put(391,205.67){\rule{1.445pt}{0.400pt}}
\multiput(391.00,206.17)(3.000,-1.000){2}{\rule{0.723pt}{0.400pt}}
\put(397,204.67){\rule{1.445pt}{0.400pt}}
\multiput(397.00,205.17)(3.000,-1.000){2}{\rule{0.723pt}{0.400pt}}
\put(416,203.67){\rule{1.445pt}{0.400pt}}
\multiput(416.00,204.17)(3.000,-1.000){2}{\rule{0.723pt}{0.400pt}}
\put(403.0,205.0){\rule[-0.200pt]{3.132pt}{0.400pt}}
\put(428,203.67){\rule{1.445pt}{0.400pt}}
\multiput(428.00,203.17)(3.000,1.000){2}{\rule{0.723pt}{0.400pt}}
\put(422.0,204.0){\rule[-0.200pt]{1.445pt}{0.400pt}}
\put(447,204.67){\rule{1.445pt}{0.400pt}}
\multiput(447.00,204.17)(3.000,1.000){2}{\rule{0.723pt}{0.400pt}}
\put(453,205.67){\rule{1.445pt}{0.400pt}}
\multiput(453.00,205.17)(3.000,1.000){2}{\rule{0.723pt}{0.400pt}}
\put(459,206.67){\rule{1.445pt}{0.400pt}}
\multiput(459.00,206.17)(3.000,1.000){2}{\rule{0.723pt}{0.400pt}}
\put(465,208.17){\rule{1.300pt}{0.400pt}}
\multiput(465.00,207.17)(3.302,2.000){2}{\rule{0.650pt}{0.400pt}}
\put(471,210.17){\rule{1.300pt}{0.400pt}}
\multiput(471.00,209.17)(3.302,2.000){2}{\rule{0.650pt}{0.400pt}}
\put(477,212.17){\rule{1.500pt}{0.400pt}}
\multiput(477.00,211.17)(3.887,2.000){2}{\rule{0.750pt}{0.400pt}}
\put(484,214.17){\rule{1.300pt}{0.400pt}}
\multiput(484.00,213.17)(3.302,2.000){2}{\rule{0.650pt}{0.400pt}}
\multiput(490.00,216.61)(1.132,0.447){3}{\rule{0.900pt}{0.108pt}}
\multiput(490.00,215.17)(4.132,3.000){2}{\rule{0.450pt}{0.400pt}}
\multiput(496.00,219.61)(1.132,0.447){3}{\rule{0.900pt}{0.108pt}}
\multiput(496.00,218.17)(4.132,3.000){2}{\rule{0.450pt}{0.400pt}}
\multiput(502.00,222.60)(0.774,0.468){5}{\rule{0.700pt}{0.113pt}}
\multiput(502.00,221.17)(4.547,4.000){2}{\rule{0.350pt}{0.400pt}}
\multiput(508.00,226.59)(0.599,0.477){7}{\rule{0.580pt}{0.115pt}}
\multiput(508.00,225.17)(4.796,5.000){2}{\rule{0.290pt}{0.400pt}}
\multiput(514.00,231.59)(0.492,0.485){11}{\rule{0.500pt}{0.117pt}}
\multiput(514.00,230.17)(5.962,7.000){2}{\rule{0.250pt}{0.400pt}}
\multiput(521.59,238.00)(0.482,1.575){9}{\rule{0.116pt}{1.300pt}}
\multiput(520.17,238.00)(6.000,15.302){2}{\rule{0.400pt}{0.650pt}}
\put(434.0,205.0){\rule[-0.200pt]{3.132pt}{0.400pt}}
\end{picture}
\vskip-1.8cm {\it Fig04: Diagrammatic representation for the identity (22).}
\end{center}

At this point we can ask ourselves: what does it mean the equations (18),
(21), and (22)? Clearly, when the explicit calculations of all the involved
amplitudes are performed, from the point of view of any chosen calculational
strategy or regularization technique, independent of the value to be
attributed by the adopted method for the involved structures, the relations
should be maintained. Note that the value for the two-point functions
involved seems to play a role of conditions for the corresponding Ward
identity preservation (the crossed channel leads to similar relations). This
does not mean that the calculation of the two-point function structures
guarantees that the symmetry properties of the three-point functions are
automatically verified, even when the desirable value is obtained for the
right hand side of the equations. It only means that the arbitrariness
involved in the explicit evaluation of the three-point functions are the
same as those we find in the evaluation of the corresponding two-point
functions. In what follows we expect to clarify all these points. In order
to give some progress in our discussions, we need to explicitly calculate
the involved structures. So, in the next section we will define our
calculational strategy to handle divergences.

\section{The Calculational Method to Handle Divergent Integrals}

As we have discussed in the previous section, if the explicit evaluation of
perturbative (divergent) amplitudes is in order, we need to specify a
philosophy to handle the mathematical indefinitions involved. Usually the
calculations become reliable only after the adoption of a regularization
technique. After this, in the intermediary steps, we invariably assume some
specific consequences for the results, intrinsically associated to the
properties attributed to the divergent integrals resulting from the
(arbitrariness) choices for the mathematical indefinitions, implied by the
adopted regularization. In the final form this way obtained for the
amplitudes in general, it is not possible to specify in a clear way what are
the particular effects of the adopted regularization for the result or, in
other words, to evaluate in what sense the expression is dependent on the
adopted technique. In order to perform, in a way as safe as possible, an
analysis of the properties of the divergent amplitudes, including their
symmetry relations and the question of ambiguities related to the
arbitrariness involved in the routing of the loop internal lines momenta, we
need to avoid as much as possible specific choices in intermediary steps in
such a way that all possibilities remain still contained in our final
results. If it is possible, we can change the usual focus of the analysis
from the verification by testing the consistency of the regularization
technique, to the identification of eventual properties that a technique
should have to be consistent. The implication of the preceding argument,
that will become clear in what follows, play a central role in the
discussions we want to make.

To explicitly evaluate the divergent integrals involved, we adopted an
alternative strategy to handle divergences \cite{Orimar-TESE}. Rather than
the specification of some regularization, to justify all the necessary
manipulations, we will assume the presence of a regulating distribution only
in an implicit way. Schematically 
\begin{equation}
\int \frac{d^{4}k}{\left( 2\pi \right) ^{4}}f(k)\rightarrow \int \frac{d^{4}k%
}{\left( 2\pi \right) ^{4}}f(k)\left\{ \lim_{\Lambda _{i}^{2}\rightarrow
\infty }G_{\Lambda _{i}}\left( k,\Lambda _{i}^{2}\right) \right\}
=\int_{\Lambda }\frac{d^{4}k}{\left( 2\pi \right) ^{4}}f(k).
\end{equation}
Here $\Lambda _{i}^{\prime }s$ are parameters of the generic distribution $%
G(\Lambda _{i}^{2},k)$ that, in addition to the obvious finiteness character
of the modified integral, must have two other very general properties; it
must be even in the integrating momentum $k$, due to Lorentz invariance
maintenance, and as well as a well-defined connection limit must exist,
i.e., 
\begin{equation}
\lim_{\Lambda _{i}^{2}\rightarrow \infty }G_{\Lambda _{i}}\left(
k^{2},\Lambda _{i}^{2}\right) =1.
\end{equation}
The first property implies that all odd integrals vanish. The second one
guarantees, in particular, that the value of the finite integrals in the
amplitudes will not be modified. Having this in mind \ we manipulate the
integrand of the divergent integrals to generate a mathematical expression
where all the divergences are located in internal momenta independent
structures. This goal can be achieved by using an adequate identity like 
\begin{equation}
\frac{1}{[(k+k_{i})^{2}-m^{2}]}=\sum_{j=0}^{N}\frac{\left( -1\right)
^{j}\left( k_{i}^{2}+2k_{i}\cdot k\right) ^{j}}{\left( k^{2}-m^{2}\right)
^{j+1}}+\frac{\left( -1\right) ^{N+1}\left( k_{i}^{2}+2k_{i}\cdot k\right)
^{N+1}}{\left( k^{2}-m^{2}\right) ^{N+1}\left[ \left( k+k_{i}\right)
^{2}-m^{2}\right] },
\end{equation}
where $k_{i}$ is (in principle) an arbitrary choice for the routing of a
loop internal line momentum. The value for $N$ should be adequately chosen.
The minor value should be the one that leads the last term in the above
expression to be present in a finite integral, and therefore, by virtue of
the well-defined connection limit assumptions, the corresponding integration
can be performed without restrictions and free from specific effects of the
eventual regularization. All the remaining structures become independent of
the internal lines momenta. We then eliminate all the integrals with odd
integrand, as a trivial consequence of the even character of the regulating
implicit distribution. In the divergent structures obtained this way, no
additional assumptions are taken. They are organized in five objects, namely 
\begin{eqnarray}
\bullet \Box _{\alpha \beta \mu \nu } &=&\int_{\Lambda }\frac{d^{4}k}{\left(
2\pi \right) ^{4}}\frac{24k_{\mu }k_{\nu }k_{\alpha }k_{\beta }}{\left(
k^{2}-m^{2}\right) ^{4}}-g_{\alpha \beta }\int_{\Lambda }\frac{d^{4}k}{%
\left( 2\pi \right) ^{4}}\frac{4k_{\mu }k_{\nu }}{\left( k^{2}-m^{2}\right)
^{3}}  \nonumber \\
&&-g_{\alpha \nu }\int_{\Lambda }\frac{d^{4}k}{\left( 2\pi \right) ^{4}}%
\frac{4k_{\beta }k_{\mu }}{\left( k^{2}-m^{2}\right) ^{3}}-g_{\alpha \mu
}\int_{\Lambda }\frac{d^{4}k}{\left( 2\pi \right) ^{4}}\frac{4k_{\beta
}k_{\nu }}{\left( k^{2}-m^{2}\right) ^{3}} \\
\bullet \Delta _{\mu \nu } &=&\int_{\Lambda }\frac{d^{4}k}{\left( 2\pi
\right) ^{4}}\frac{4k_{\mu }k_{\nu }}{\left( k^{2}-m^{2}\right) ^{3}}%
-\int_{\Lambda }\frac{d^{4}k}{\left( 2\pi \right) ^{4}}\frac{g_{\mu \nu }}{%
\left( k^{2}-m^{2}\right) ^{2}} \\
\bullet \nabla _{\mu \nu } &=&\int_{\Lambda }\frac{d^{4}k}{\left( 2\pi
\right) ^{4}}\frac{2k_{\nu }k_{\mu }}{\left( k^{2}-m^{2}\right) ^{2}}%
-\int_{\Lambda }\frac{d^{4}k}{\left( 2\pi \right) ^{4}}\frac{g_{\mu \nu }}{%
\left( k^{2}-m^{2}\right) } \\
\bullet I_{log}(m^{2}) &=&\int_{\Lambda }\frac{d^{4}k}{\left( 2\pi \right)
^{4}}\frac{1}{\left( k^{2}-m^{2}\right) ^{2}} \\
\bullet I_{quad}(m^{2}) &=&\int_{\Lambda }\frac{d^{4}k}{\left( 2\pi \right)
^{4}}\frac{1}{\left( k^{2}-m^{2}\right) }.
\end{eqnarray}
This systematization is sufficient for discussions in fundamental theories
at the one-loop level \cite{Orimar-NPB}. In non-renormalizable ones, new
objects can be defined following this philosophy \cite
{Orimar-PRD,Orimar-Mota}. In the two (or more) loop level of calculations
new basic divergent structures can be equally defined in a completely
analogous way. The main point is to avoid the explicit evaluation of such
divergent structures, in which case a regulating distribution needs to be
specified.

We can say that this procedure furnishes an universal point of view for the
calculated amplitudes, once it becomes possible to map the final expressions
obtained this way into the corresponding results of other techniques. All
the steps followed and all the assumptions are perfectly valid in the
reasonable regularization prescriptions, including the DR. All we need, to
extract from our results those of a specific technique, is to evaluate the
divergent structures remaining at the final expression according to the
specific chosen prescription. Another important fact we call the attention
is that no shifts or expansions are used in intermediary steps. We assume
the most general as possible routing for all amplitudes. The potential
ambiguous terms are still present in the final results. Consequently it is
possible to make contact with those corresponding to the explicit evaluation
of surface term involved when shifts in the internal momenta are performed.
This is an important aspect of our analysis because we want to make contact
with the traditional approach used to justify the triangle anomalies.

In order to clarify the above described method, to handle divergences, let
us apply the calculational strategy in the treatment of some divergent
integrals. For this purpose we take two of them that will play an important
role in our analysis. They are two-point functions structures defined as
follows 
\begin{equation}
\left( I_{2};I_{2}^{\mu }\right) =\int \frac{d^{4}k}{\left( 2\pi \right) ^{4}%
}\frac{\left( 1;k^{\mu }\right) }{\left[ \left( k+k_{1}\right) ^{2}-m^{2}%
\right] \left[ \left( k+k_{2}\right) ^{2}-m^{2}\right] }.
\end{equation}
The first indicated above, the $I_{2}$ integral is a logarithmically
divergent structure while $\left( I_{2}\right) _{\mu }$ is a linearly one.
In these structures $k_{1}$ and $k_{2}$ represent, in principle, arbitrary
choices for the internal lines momenta. Therefore we can expect a dependence
on $k_{1}$ and $k_{2}$ other than the difference between them only for the $%
\left( I_{2}\right) _{\mu }$ integral.

Taken first the $I_{2}$ integral we choose, in the identity (25), $N=1$ to
rewrite both denominators. Then we get 
\begin{eqnarray}
I_{2} &=&\int_{\Lambda }\frac{d^{4}k}{(2\pi )^{4}}\frac{1}{(k^{2}-m^{2})^{2}}
\nonumber \\
&&-\int \frac{d^{4}k}{(2\pi )^{4}}\frac{(k_{1}^{2}+2k_{1}\cdot k)^{2}}{%
(k^{2}-m^{2})^{2}[(k+k_{1})^{2}-m^{2}]}  \nonumber \\
&&-\int \frac{d^{4}k}{(2\pi )^{4}}\frac{(k_{2}^{2}+2k_{2}\cdot k)^{2}}{%
(k^{2}-m^{2})^{2}[(k+k_{2})^{2}-m^{2}]}  \nonumber \\
&&+\int \frac{d^{4}k}{(2\pi )^{4}}\frac{(k_{1}^{2}+2k_{1}\cdot
k)(k_{2}^{2}+2k_{2}\cdot k)}{%
(k^{2}-m^{2})^{2}[(k+k_{1})^{2}-m^{2}][(k+k_{2})^{2}-m^{2}]}.
\end{eqnarray}
The right hand side exhibits the desirable form. The divergent term is
located in a structure which is independent of internal momenta and which we
can identify as $I_{\log }\left( m^{2}\right) ,$ defined in equation (29).
The remaining structures are finite ones and we use what we call connection
limit existence to drop the $\Lambda $ subscript on the integral, or
equivalently to remove the eventual regulating distribution under the
argumentation that the integration and the connection limit can be perfectly
interchanged. The thus obtained finite Feynman integrals can be solved
without any problem. The answer can be written as 
\begin{equation}
I_{2}=I_{log}(m^{2})-\left( \frac{i}{(4\pi )^{2}}\right)
Z_{0}((k_{1}-k_{2})^{2};m^{2}),
\end{equation}
where we have introduced (in shorthand notation) the two-point function
structures 
\begin{equation}
Z_{k}(\lambda _{1}^{2},\lambda _{2}^{2},q^{2};\lambda
^{2})=\int_{0}^{1}dzz^{k}ln\left( \frac{q^{2}z(1-z)+(\lambda
_{1}^{2}-\lambda _{2}^{2})z-\lambda _{1}^{2}}{(-\lambda ^{2})}\right) .
\end{equation}
Analytical expressions can be easily obtained \cite{Orimar-TESE} but for our
present purposes this is not necessary.

Following the procedure we can also evaluate the $I_{2}^{\mu }$ integral.
The first step is the same: the use of the identity (25) to rewrite the
integrand, now to the form 
\begin{eqnarray}
\left( I_{2}\right) _{\mu } &=&-\frac{1}{2}(k_{1}+k_{2})_{\alpha
}\int_{\Lambda }\frac{d^{4}k}{(2\pi )^{4}}\frac{4k_{\alpha }k_{\mu }}{%
(k^{2}-m^{2})^{3}}  \nonumber \\
&&+\int \frac{d^{4}k}{(2\pi )^{4}}\frac{(k_{1}^{2}+2k_{1}\cdot k)^{2}k_{\mu }%
}{(k^{2}-m^{2})^{3}[(k+k_{1})^{2}-m^{2}]}  \nonumber \\
&&+\int \frac{d^{4}k}{(2\pi )^{4}}\frac{(k_{2}^{2}+2k_{2}\cdot k)^{2}k_{\mu }%
}{(k^{2}-m^{2})^{3}[(k+k_{2})^{2}-m^{2}]}  \nonumber \\
\!\!\!\!\!\!\!\!\!\!\!\!\!\!\!\!\!\!\!\!\!\!\! &&+\int \frac{d^{4}k}{(2\pi
)^{4}}\frac{(k_{1}^{2}+2k_{1}\cdot k)(k_{2}^{2}+2k_{2}\cdot k)k_{\mu }}{%
(k^{2}-m^{2})^{2}[(k+k_{1})^{2}-m^{2}][(k+k_{2})^{2}-m^{2}]}.
\end{eqnarray}
In the above expression, we have dropped two odd $k$ integrals, by virtue of
the even character of the implicit regulating distribution as well as the $%
\Lambda $ subscript in the last three terms due to the finite character.
After the integration of the finite terms, we are led to the result 
\begin{eqnarray}
\left( I_{2}\right) _{\mu } &=&-\frac{1}{2}(k_{1}+k_{2})_{\alpha }\left(
\Delta _{\alpha \mu }\right) \!-\frac{1}{2}(k_{1}+k_{2})_{\mu }\left\{
I_{\log }(m^{2})-\left( \frac{i}{(4\pi )^{2}}\right)
Z_{0}((k_{1}-k_{2})^{2};m^{2})\right\}  \nonumber \\
&=&-\frac{1}{2}(k_{1}+k_{2})_{\alpha }\left( \Delta _{\alpha \mu }\right) -%
\frac{1}{2}(k_{1}+k_{2})_{\mu }\left( I_{2}\right) .
\end{eqnarray}
Following the same prescription, for future use, we can calculate three new
integrals; 
\begin{equation}
\left( I_{3};I_{3}^{\mu };I_{3}^{\mu \nu }\right) =\int \frac{d^{4}k}{(2\pi
)^{4}}\frac{\left( 1;k^{\mu };k^{\mu }k^{\nu }\right) }{%
[(k+k_{1})^{2}-m^{2}][(k+k_{2})^{2}-m^{2}]\left[ (k+k_{3})^{2}-m^{2}\right] }%
.
\end{equation}
This is a very easy job because only one of them is a divergent structure.
We write the results as 
\begin{eqnarray}
&&\bullet I_{3}=\left( \frac{i}{(4\pi )^{2}}\right) \xi _{00} \\
&&\bullet \left( I_{3}\right) _{\mu }=\left( \frac{i}{(4\pi )^{2}}\right)
\{\left( k_{1}-k_{2}\right) _{\mu }\xi _{01}-\left( k_{3}-k_{1}\right) _{\mu
}\xi _{10}\}-k_{1\mu }I_{3} \\
&&\bullet \left( I_{3}\right) _{\mu \nu }=\left( \frac{i}{(4\pi )^{2}}%
\right) \left\{ -\frac{g_{\mu \nu }}{2}\left[ \eta _{00}\right] \;+\left(
k_{1}-k_{2}\right) _{\mu }\left( k_{1}-k_{2}\right) _{\nu }\xi _{02}+\left(
k_{3}-k_{1}\right) _{\mu }\left( k_{3}-k_{1}\right) _{\nu }\xi _{20}\right. 
\nonumber \\
&&\;\;\;\;\;\;\left. -\left( k_{1}-k_{2}\right) _{\mu }\left(
k_{3}-k_{1}\right) _{\nu }\xi _{11}-\left( k_{1}-k_{2}\right) _{\nu }\left(
k_{3}-k_{1}\right) _{\mu }\xi _{11}\right\}  \nonumber \\
&&\;\;\;\;\;\;+\frac{g_{\mu \nu }}{4}\left[ I_{\log }\left( m^{2}\right) %
\right] +\frac{\Delta _{\mu \nu }}{4}-k_{1\mu }\left( I_{3}\right) _{\nu
}-k_{1\nu }\left( I_{3}\right) _{\mu }+k_{1\nu }k_{1\mu }I_{3},
\end{eqnarray}
where we have introduced the three-point function structures ${\xi }_{nm}$
defined as 
\begin{equation}
\xi _{nm}(k_{3}-k_{1},k_{1}-k_{2};m)=\int_{0}^{1}\,dz\int_{0}^{1-z}\,dy{%
\frac{z^{n}y^{m}}{Q(y,z)}},
\end{equation}
where $Q(y,z)=\left( k_{1}-k_{2}\right) ^{2}y(1-y)+\left( k_{3}-k_{1}\right)
^{2}z(1-z)+2\left( k_{1}-k_{2}\right) \cdot \left( k_{3}-k_{1}\right)
yz-m^{2},$ and 
\begin{equation}
\eta _{00}={\frac{1}{2}}Z_{0}((k_{3}-k_{2})^{2};m^{2})-\left( {\frac{1}{2}}%
+m^{2}\xi _{00})\right) +{\frac{1}{2}}\left( k_{3}-k_{1}\right) ^{2}\xi
_{10}+{\frac{1}{2}}\left( k_{1}-k_{2}\right) ^{2}\xi _{01}.
\end{equation}
At this point it is important to emphasize the general aspects of the
method. No shifts have been performed and, in fact, no divergent integrals
have been calculated. All final results produced by this approach can be
mapped into those of any specific technique. The finite parts are the same
as they should be by physical reasons. The divergent parts can be easily
obtained. All we need is to evaluate the remaining divergent structures. By
virtue of this general character, the present strategy can be simply used to
systematize the procedures, even if one wants to use traditional techniques.
Those parts that depend on the specific regularization method are naturally
separated allowing us to analyze such dependence in a particular problem,
which is very interesting separately. Let us now use the above obtained
result to calculate physical amplitudes.

\section{Two-Point Functions evaluation and the Traditional Way to Look at
Anomalies}

With the calculational strategy presented in the previous section, the
evaluation of physical amplitudes becomes a direct and simple job. First we
perform the involved traces in order to write the amplitudes as a
combination of Feynman integral even if some of them are divergent
mathematical structures. After the identification of the integrals, we
simply bring the results previously obtained. Let us take, as a first
example, the two-point function structure obtained in the right hand side of
the equations (18), (21), and (22). After the traces have been performed we
write, 
\begin{equation}
T_{\mu \nu }^{AV}(k_{1},k_{2};m)=2\varepsilon _{\mu \nu \alpha \beta
}(k_{1}-k_{2})_{\beta }\left\{ (k_{1}+k_{2})_{\alpha }I_{2}+2\left(
I_{2}\right) _{\alpha }\right\} .
\end{equation}
The two integrals we need have already been calculated. Substituting the
expressions (33) and (36) we get: 
\begin{equation}
T_{\mu \nu }^{AV}=2\varepsilon _{\mu \nu \alpha \beta }(k_{1}-k_{2})_{\alpha
}(k_{1}+k_{2})_{\xi }\left[ \triangle _{\xi \beta }\right] .
\end{equation}
The above expression represents the last step we can perform without
assuming any arbitrary choices, in order to get a definite result for the
physical amplitude. First, we need to adopt some regularization recipe or
equivalent philosophy to attribute some significance to the object $%
\triangle $, once it is an undefined mathematical structure. Such a choice
is arbitrary because it is not present in the Feynman rules of a theory or
model that has generated the $AV$ amplitude. If the result attributed to the 
$\triangle $ object presents a dependence on the specific regularization
adopted, then the arbitrariness becomes an ambiguity.

Another arbitrariness, which plays an important role in such type of
discussions, is that related to the choices for the routing of the internal
lines momenta. In our present problem the choices are taken as the most
general ones. Only the difference $k_{1}-k_{2}$ is a definite physical
quantity, in such a way the result (44) presents a potentially ambiguous
character. For this purpose it is sufficient that the object $\triangle $
assumes a nonzero value as a consequence of the adopted regularization. This
is materialized by the dependence on $k_{1}+k_{2}$. Again, we are free to
choose $k_{1}$ and $k_{2}$, this is arbitrary, but if our final result
presents a dependence on this choice, which seems to be this case, then the
arbitrariness will become an ambiguity.

Let us now return to the question of the symmetry relations for the $AVV$
amplitude. Inserting the results for the $AV$ term we obtain for the
equations (18), (21), and (22) respectively 
\begin{eqnarray}
\bullet (k_{3}-k_{2})_{\lambda }T_{\lambda \mu \nu }^{AVV} &=&-2mi[T_{\mu
\nu }^{PVV}]  \nonumber \\
&&+2\varepsilon _{\mu \nu \alpha \beta }\left[ (k_{1}-k_{3})_{\beta
}(k_{1}+k_{3})_{\xi }+(k_{2}-k_{1})_{\beta }(k_{1}+k_{2})_{\xi }\right]
\triangle _{\xi \alpha } \\
\bullet (k_{3}-k_{1})_{\mu }T_{\lambda \mu \nu }^{AVV} &=&2\varepsilon
_{\lambda \nu \alpha \beta }\left[ (k_{2}-k_{1})_{\beta }(k_{1}+k_{2})_{\xi
}+(k_{3}-k_{2})_{\beta }(k_{2}+k_{3})_{\xi }\right] \triangle _{\xi \alpha }
\\
\bullet (k_{1}-k_{2})_{\nu }T_{\lambda \mu \nu }^{AVV} &=&2\varepsilon
_{\lambda \mu \alpha \beta }\left[ (k_{3}-k_{1})_{\beta }(k_{1}+k_{3})_{\xi
}+(k_{2}-k_{3})_{\beta }(k_{2}+k_{3})_{\xi }\right] \triangle _{\xi \alpha }.
\end{eqnarray}
Before the analysis we want to make from our proposed point of view, let us
show how the general result put above can furnish what we call the
traditional way to look at the anomalies. The referred point of view assumes
that the resulting amplitudes may be ambiguous quantities and after the
ambiguous term has been identified, performing a shift on the integrating
momentum, the corresponding surface term is evaluated. In the ambiguous
terms remaining, we need to make a choice in such a way to attribute
definite significance for the amplitudes. Such a choice is made following
symmetry arguments. Due to the fact that in order to obtain our result no
shifts have been performed and no specific assumptions for the divergent
terms are introduced, our expressions can be converted into that obtained by
the traditional procedure. For this purpose it is sufficient to note the
identity 
\begin{equation}
\int_{\Lambda }\frac{d^{4}k}{\left( 2\pi \right) ^{4}}\frac{\partial }{%
\partial k_{\nu }}\left( \frac{k_{\mu }}{\left( k^{2}-m^{2}\right) ^{2}}%
\right) =\int_{\Lambda }\frac{d^{4}k}{\left( 2\pi \right) ^{4}}\frac{%
-4k_{\mu }k_{\nu }}{\left( k^{2}-m^{2}\right) ^{3}}+\int_{\Lambda }\frac{%
d^{4}k}{\left( 2\pi \right) ^{4}}\frac{g_{\mu \nu }}{\left(
k^{2}-m^{2}\right) ^{2}}=-\Delta _{\mu \nu }.
\end{equation}
The left hand side is a total derivative which, due to the Gauss theorem, is
a surface term. The value can be immediately obtained as 
\begin{equation}
\triangle _{\mu \nu }=-\left( \frac{i}{32\pi ^{2}}\right) g_{\mu \nu },
\end{equation}
which is then identified as the surface term evaluated in the traditional
approach \cite{Jackiw1,Jackiw2}. In the next step, once we have a dependence
on arbitrary momenta given this value for $\triangle $, we parameterize the
internal momenta in terms of the external ones. We adopt then 
\begin{equation}
\left\{ 
\begin{array}{l}
k_{1}=ap^{\prime }+bp \\ 
k_{2}=bp+(a-1)p^{\prime } \\ 
k_{3}=ap^{\prime }+(b+1)p,
\end{array}
\right.
\end{equation}
where $a$ and $b$ are constants. Notice that : $k_{1}-k_{2}=p^{\prime
},\;k_{3}-k_{1}=p$ and $k_{3}-k_{2}=p^{\prime }+p=q$, where $q$ is obviously
the momentum of the axial vector. After these substitutions we get 
\begin{eqnarray}
\bullet \left( k_{3}-k_{2}\right) _{\lambda }T_{\lambda \mu \nu }^{AVV}
&=&-2miT_{\mu \nu }^{PVV}-\frac{\left( a-b\right) }{8\pi ^{2}}i\varepsilon
_{\mu \nu \alpha \beta }p^{\alpha }p^{\prime \beta } \\
\bullet \left( k_{3}-k_{1}\right) _{\mu }T_{\lambda \mu \nu }^{AVV} &=&-%
\frac{\left( 1-a\right) }{8\pi ^{2}}i\varepsilon _{\lambda \nu \alpha \beta
}p^{\alpha }p^{\prime \beta } \\
\bullet \left( k_{1}-k_{2}\right) _{\nu }T_{\lambda \mu \nu }^{AVV} &=&\frac{%
\left( 1+b\right) }{8\pi ^{2}}i\varepsilon _{\lambda \mu \alpha \beta
}p^{\alpha }p^{\prime \beta }.
\end{eqnarray}

In the above expressions the arbitrariness related to the routing of
internal lines now present in the parameters $a$ and $b$ remains. In
addition we note that there are no values for $a$ and $b$ in such a way that
all the expected relations among the involved Green's functions are
simultaneously satisfied. If we follow this line of reasoning and include
the contribution of the crossed diagram whose parameterization for the
internal lines momenta can be assumed as 
\begin{equation}
\left\{ 
\begin{array}{l}
l_{1}=cp+dp^{\prime } \\ 
l_{2}=dp^{\prime }+(c-1)p \\ 
l_{3}=cp+(d+1)p^{\prime },
\end{array}
\right.
\end{equation}
we will obtain 
\begin{eqnarray}
\bullet (l_{3}-l_{2})_{\lambda }T_{\lambda \nu \mu }^{AVV} &=&-2miT_{\nu \mu
}^{PVV}-\frac{\left( c-d\right) }{8\pi ^{2}}i\varepsilon _{\mu \nu \alpha
\beta }p_{\alpha }p_{\beta }^{\prime } \\
\bullet (l_{1}-l_{2})_{\mu }T_{\lambda \nu \mu }^{AVV} &=&-\frac{\left(
d+1\right) }{8\pi ^{2}}i\varepsilon _{\lambda \nu \alpha \beta }p_{\alpha
}p_{\beta }^{\prime } \\
\bullet (l_{3}-l_{1})_{\nu }T_{\lambda \nu \mu }^{AVV} &=&-\frac{c-1}{8\pi
^{2}}i\varepsilon _{\lambda \mu \alpha \beta }p_{\alpha }p_{\beta }^{\prime
}.
\end{eqnarray}
The addition of the two contributions gives us 
\begin{eqnarray}
\bullet q_{\lambda }T_{\lambda \mu \nu }^{A\rightarrow VV} &=&-2miT_{\mu \nu
}^{P\rightarrow VV}-\frac{\left( a-b+c-d\right) }{8\pi ^{2}}i\varepsilon
_{\mu \nu \alpha \beta }p_{\alpha }p_{\beta }^{\prime } \\
\bullet p_{\mu }T_{\lambda \mu \nu }^{A\rightarrow VV} &=&-\frac{\left(
d-a+2\right) }{8\pi ^{2}}i\varepsilon _{\lambda \nu \alpha \beta }p_{\alpha
}p_{\beta }^{\prime } \\
\bullet p_{\nu }^{\prime }T_{\lambda \mu \nu }^{A\rightarrow VV} &=&-\frac{%
\left( c-b-2\right) }{8\pi ^{2}}i\varepsilon _{\lambda \mu \alpha \beta
}p_{\alpha }p_{\beta }^{\prime }.
\end{eqnarray}
A closer contact with the usual results can be obtained if it is assumed the
same significance for the arbitrary internal momenta, i.e., $a=c$ and $b=d$
in the equations (51)-(53), and (55)-(57). We get then 
\begin{eqnarray}
\bullet q_{\lambda }T_{\lambda \mu \nu }^{A\rightarrow VV} &=&-2miT_{\mu \nu
}^{P\rightarrow VV}-\frac{\left( a-b\right) }{4\pi ^{2}}i\varepsilon _{\mu
\nu \alpha \beta }p_{\alpha }p_{\beta }^{\prime } \\
\bullet p_{\mu }T_{\lambda \mu \nu }^{A\rightarrow VV} &=&-\frac{\left(
b-a+2\right) }{8\pi ^{2}}i\varepsilon _{\lambda \nu \alpha \beta }p_{\alpha
}p_{\beta }^{\prime } \\
\bullet p_{\nu }^{\prime }T_{\lambda \mu \nu }^{A\rightarrow VV} &=&-\frac{%
\left( a-b-2\right) }{8\pi ^{2}}i\varepsilon _{\lambda \mu \alpha \beta
}p_{\alpha }p_{\beta }^{\prime }.
\end{eqnarray}
Finally, we choose the value $a=1$ in the above expression to get 
\begin{eqnarray}
\bullet q_{\lambda }T_{\lambda \mu \nu }^{A\rightarrow VV} &=&-2miT_{\mu \nu
}^{P\rightarrow VV}-\frac{\left( 1-b\right) }{4\pi ^{2}}i\varepsilon _{\mu
\nu \alpha \beta }p_{\alpha }p_{\beta }^{\prime } \\
\bullet p_{\mu }T_{\lambda \mu \nu }^{A\rightarrow VV} &=&-\frac{\left(
1+b\right) }{8\pi ^{2}}i\varepsilon _{\lambda \nu \alpha \beta }p_{\alpha
}p_{\beta }^{\prime } \\
\bullet p_{\nu }^{\prime }T_{\lambda \mu \nu }^{A\rightarrow VV} &=&\frac{%
\left( 1+b\right) }{8\pi ^{2}}i\varepsilon _{\lambda \mu \alpha \beta
}p_{\alpha }p_{\beta }^{\prime }.
\end{eqnarray}
The result this way obtained, can be immediately recognized as the
traditional one \cite{Jackiw1}\cite{Livros}\cite{Cheng-Li}. It is now clear
that there is no value for the $b$ parameter in order to preserve all Ward
identities. Following the usual arguments and choosing the value $b=-1$ the $%
U(1)$ gauge symmetry is maintained, the axial one is violated. However, this
analysis cannot reveal us anything about the fourth symmetry property of the 
$AVV$ physical amplitude, which is certainly the most important aspect
related to the fundamental nature of the anomaly phenomenon. The low-energy
limit of the $AVV$ amplitude is modified because of the presence of the
anomalous term and the pion decay rate is obtained in agreement with the
experimental data. This result is compatible with the statements of the
Sutherland-Veltman paradox: two symmetry relations and the low energy
theorem are chosen to be satisfied. One symmetry relation is assumed
violated. If we stop our discussion at this point, nothing new is added to
the status of the problem. All our presentations can be considered as an
alternative way to perform the calculations, perhaps more simple and direct
but nothing more than this.

Let us now analyze the equations (45), (46), and (47), which are completely
arbitrary, but now following our arguments. We call the attention to the
fact that the $AV$ amplitude, which usually is looked only as an integral,
is in fact a physical amplitude. There are, therefore, physical consequences
implied by the assumed value for such amplitude. All the possibilities for
the choices of the undefined quantities should be considered in light of the
consistency requirements imposed by physical reasons. So, we then ask for
the physical constraints that the $AV$ is submitted. First, it is a
two-point function with two internal propagators. Due to unitarity reasons
(Cutkosky's rules) it should exhibit a complex threshold at the external
momentum $\left( k_{1}-k_{2}\right) ^{2}=4m^{2}$. Looking at the expression
(44) such possibility is not available by the arbitrariness still present.
So, if it is not vanishing, it is not compatible with the unitarity. For the
second, if a nonzero value is attributed, the amplitude may connect an axial
particle to a vector one. CPT is clearly violated. We can add other symmetry
reasons stated by Ward identities. The $AV$ amplitude possesses two Lorentz
indexes and therefore there are two Ward identities. The contractions with
the external momentum reveal 
\begin{equation}
\left( k_{1}-k_{2}\right) _{\mu }T_{\mu \nu }^{AV}=\left( k_{1}-k_{2}\right)
_{\nu }T_{\mu \nu }^{AV}=0.
\end{equation}

If $T_{\mu \nu }^{AV}$ is not identically zero it carries two conserved
currents, which is clearly inconsistent. With the above argumentation it is
clear that there are many reasons to constrain the $AV$ to the identically
zero value. Guided by these reasons we ask for the ways that we have to make
choices in order to obtain this value. There are two very different options.
The first is to choose a regularization such that $\Delta _{\mu \nu }^{reg}$ 
$=0.$ With this choice we automatically eliminate the potentially ambiguous
terms. The second option is to choose the ambiguous dependence on internal
momenta in such a way $k_{1}+k_{2}=0.$ \ In this case it is immaterial the
value for $\Delta _{\mu \nu }.$ There is a price to be paid in adopting such
a choice. In all the amplitudes in all theories or models the procedure
should be the same, if we want a method. If one assumes the case-by-case
analysis, i.e., in one problem a zero value for the $\Delta _{\mu \nu }$ is
assumed by some reason and in another a different value may be accepted,
then our worries do not make sense. The advantage of the choice $\Delta
_{\mu \nu }^{reg}=0$ is that it is a property for divergent integrals, not
for the amplitudes. In the DR technique such a value is automatically zero
where this method can be applied. If we choose the value for the arbitrary
momenta, then we are assuming that the physical amplitudes are intrinsically
ambiguous quantities. The predictive power of the QFT in perturbative
solutions is completely destroyed. We can use it only for adjustments of
well-known phenomenologies. Predictions cannot be made in a definite way. In
addition, we are assuming that the space-time homogeneity is lost in the
calculations. This is evidently unacceptable as a final form of knowledge.
So, it seems very reasonable to assume that the only consistent possibility
is the choice $\Delta _{\mu \nu }^{reg}=0$ and we ask then for the
consequences. First we note that the symmetry relations (45), (46) and (47)
will lead us to 
\begin{eqnarray}
\bullet p_{\mu }T_{\lambda \mu \nu }^{A\rightarrow VV} &=&p_{\nu }^{\prime
}T_{\lambda \mu \nu }^{A\rightarrow VV}=0 \\
\bullet q_{\lambda }T_{\lambda \mu \nu }^{A\rightarrow VV} &=&-2miT_{\mu \nu
}^{P\rightarrow VV}.
\end{eqnarray}
At first sight we can learn that this means that no symmetry relation can be
violated. It is not necessarily true but if it would be the case this would
not be inconsistent with the Sutherland-Veltman paradox, because such a
expression for the $T_{\lambda \mu \nu }^{AVV}$ should violate the low
energy theorem. The point is that, until this point, we have not explicitly
calculated the three-point functions yet. So, in particular, we have not
accessed the low energy behavior in order to verify such property. To
clarify these statements, let us consider in the next section the explicit
calculation for the $AVV$ and $PVV$ amplitudes and verify the Ward
identities by using these expressions.

\section{ Explicit Calculations for the $AVV$ and $PVV$ Amplitudes}

Let us follow the procedure we have described in the section III to
explicitly calculate the $AVV$ and $PVV$ amplitudes. First we consider the
expression (13) and after performing the Dirac traces we write 
\begin{equation}
T_{\lambda \mu \nu }^{AVV}=-4\left\{ -F_{\lambda \mu \nu }+N_{\lambda \mu
\nu }+M_{\lambda \mu \nu }+P_{\lambda \mu \nu }\right\} ,
\end{equation}
by reasons that we will see in what follows. In the above expression we have
adopted the definitions 
\begin{eqnarray}
\bullet P_{\lambda \mu \nu } &=&g_{\mu \nu }\varepsilon _{\alpha \beta
\lambda \xi }\int \frac{d^{4}k}{(2\pi )^{4}}\frac{(k+k_{1})_{\alpha
}(k+k_{2})_{\beta }(k+k_{3})_{\xi }}{%
[(k+k_{1})^{2}-m^{2}][(k+k_{2})^{2}-m^{2}][(k+k_{3})^{2}-m^{2}]} \\
\bullet F_{\lambda \mu \nu } &=&\int \frac{d^{4}k}{\left( 2\pi \right) ^{4}}%
\{\varepsilon _{\nu \beta \lambda \xi }\left( k+k_{1}\right) _{\mu }\left(
k+k_{2}\right) _{\beta }\left( k+k_{3}\right) _{\xi }  \nonumber \\
&&\;\;\;\;\;\;\;\;\;\;\;\;+\varepsilon _{\mu \beta \lambda \xi }\left(
k+k_{1}\right) _{\nu }\left( k+k_{2}\right) _{\beta }\left( k+k_{3}\right)
_{\xi }  \nonumber \\
&&\;\;\;\;\;\;\;\;\;\;\;\;+\varepsilon _{\mu \alpha \nu \beta }\left(
k+k_{1}\right) _{\alpha }\left( k+k_{2}\right) _{\beta }\left(
k+k_{3}\right) _{\lambda }  \nonumber \\
&&\;\;\;\;\;\;\;\;\;\;\;\;+\left. \varepsilon _{\mu \alpha \nu \xi }\left(
k+k_{1}\right) _{\alpha }\left( k+k_{3}\right) _{\xi }\left( k+k_{2}\right)
_{\lambda }\right\} \times  \nonumber \\
&&\;\;\;\;\;\;\;\;\;\;\;\times \left\{ \frac{1}{\left[ \left( k+k_{1}\right)
^{2}-m^{2}\right] \left[ \left( k+k_{2}\right) ^{2}-m^{2}\right] \left[
\left( k+k_{3}\right) ^{2}-m^{2}\right] }\right\} \\
\bullet N_{\lambda \mu \nu } &=&\frac{\varepsilon _{\mu \alpha \nu \lambda }%
}{2}\left\{ \int \frac{d^{4}k}{\left( 2\pi \right) ^{4}}\frac{%
(k+k_{1})_{\alpha }}{\left[ (k+k_{2})^{2}-m^{2}\right] \left[
(k+k_{1})^{2}-m^{2}\right] }\right.  \nonumber \\
&&\;\;\;\;\;\;\;+\int \frac{d^{4}k}{\left( 2\pi \right) ^{4}}\frac{%
(k+k_{1})_{\alpha }}{\left[ (k+k_{1})^{2}-m^{2}\right] \left[
(k+k_{3})^{2}-m^{2}\right] }  \nonumber \\
&&\;\;\;\;\;\;\ +[2m^{2}-(k_{2}-k_{3})^{2}]\times  \nonumber \\
&&\;\;\;\;\;\;\;\;\times \left. \int \frac{d^{4}k}{\left( 2\pi \right) ^{4}}%
\frac{(k+k_{1})_{\alpha }}{\left[ (k+k_{1})^{2}-m^{2}\right] \left[
(k+k_{2})^{2}-m^{2}\right] \left[ (k+k_{3})^{2}-m^{2}\right] }\right\} \\
\bullet M_{\lambda \mu \nu } &=&m^{2}\varepsilon _{\mu \nu \alpha \lambda
}\int \frac{d^{4}k}{\left( 2\pi \right) ^{4}}\frac{\left\{ \left(
k+k_{2}\right) _{\alpha }-\left( k+k_{1}\right) _{\alpha }+\left(
k+k_{3}\right) _{\alpha }\right\} }{\left[ \left( k+k_{1}\right) ^{2}-m^{2}%
\right] \left[ \left( k+k_{2}\right) ^{2}-m^{2}\right] \left[ \left(
k+k_{3}\right) ^{2}-m^{2}\right] }.
\end{eqnarray}
To arrive at these results only the traces were evaluated and the identity 
\begin{equation}
\left( k+k_{i}\right) \cdot \left( k+k_{j}\right) =\frac{1}{2}\left[ \left(
k+k_{i}\right) ^{2}-m^{2}\right] +\frac{1}{2}\left[ \left( k+k_{j}\right)
^{2}-m^{2}\right] +\frac{1}{2}\left[ 2m^{2}-\left( k_{i}-k_{j}\right) ^{2}%
\right] ,
\end{equation}
was used in the $N_{{\lambda }{\mu }{\nu }}$ term. Note that $F_{{\lambda }{%
\mu }{\nu }}$ is logarithmically divergent, $N_{{\lambda }{\mu }{\nu }}$ is
linearly divergent and $M_{{\lambda }{\mu }{\nu }}$ is finite. All we need
is to substitute the results for the integrals (33), (36), (38), (39), and
(40) in the expressions for the structures we have defined. We get then 
\begin{eqnarray}
\bullet P_{\lambda \mu \nu } &=&0 \\
\bullet M_{\lambda \mu \nu } &=&\left( \frac{i}{(4\pi )^{2}}\right)
\varepsilon _{\mu \alpha \nu \lambda }m^{2}\left\{ \left( k_{1}-k_{2}\right)
_{\alpha }\left( \xi _{00}-\xi _{01}\right) -\left( k_{3}-k_{1}\right)
_{\alpha }\left( \xi _{00}-\xi _{10}\right) \right\} \\
\bullet N_{\lambda \mu \nu } &=&\frac{\varepsilon _{\mu \alpha \nu \lambda }%
}{4}\left( k_{1}-k_{2}\right) _{\alpha }\left\{ I_{\log }\left( m^{2}\right)
-\left( \frac{i}{\left( 4\pi \right) ^{2}}\right) Z_{0}\left( \left(
k_{1}-k_{2}\right) ^{2};m^{2}\right) \right.  \nonumber \\
&&\;\;\;\;\;\;\;\;\;\;\;\;\;\;\;\;\;\;\;\;\;\;\;\;\left. +\left( \frac{i}{%
\left( 4\pi \right) ^{2}}\right) \left[ 2m^{2}-\left( k_{3}-k_{2}\right) ^{2}%
\right] \left( 2\xi _{01}\right) \right\}  \nonumber \\
&&-\frac{\varepsilon _{\mu \alpha \nu \lambda }}{4}\left( k_{3}-k_{1}\right)
_{\alpha }\left\{ I_{\log }\left( m^{2}\right) -\left( \frac{i}{\left( 4\pi
\right) ^{2}}\right) Z_{0}\left( \left( k_{1}-k_{3}\right) ^{2};m^{2}\right)
\right.  \nonumber \\
&&\;\;\;\;\;\;\;\;\;\;\;\;\;\;\;\;\;\;\;\;\;\;\;\;\left. +\left( \frac{i}{%
\left( 4\pi \right) ^{2}}\right) \left[ 2m^{2}-\left( k_{3}-k_{2}\right) ^{2}%
\right] \left( 2\xi _{10}\right) \right\}  \nonumber \\
&&-\frac{\varepsilon _{\mu \alpha \nu \lambda }}{4}\left[ \left(
k_{1}+k_{2}\right) _{\beta }+\left( k_{3}+k_{1}\right) _{\beta }\right]
\Delta _{\alpha \beta } \\
\bullet F_{\lambda \mu \nu } &=&\left( \frac{i}{\left( 4\pi \right) ^{2}}%
\right) \left( k_{3}-k_{1}\right) _{\xi }\left( k_{1}-k_{2}\right) _{\beta
}\left\{ \varepsilon _{\nu \beta \lambda \xi }[\left( k_{1}-k_{2}\right)
_{\mu }\left( \xi _{02}+\xi _{11}-\xi _{01}\right) \right.  \nonumber \\
&&\hspace{1in}\hspace{1in}\hspace{0.4in}-\left( k_{3}-k_{1}\right) _{\mu
}\left( \xi _{20}+\xi _{11}-\xi _{10}\right) ]  \nonumber \\
&&\hspace{1in}\hspace{1in}\hspace{0.1in}+\varepsilon _{\mu \beta \lambda \xi
}[\left( k_{1}-k_{2}\right) _{\nu }\left( \xi _{02}+\xi _{11}-\xi
_{01}\right)  \nonumber \\
&&\hspace{1in}\hspace{1in}\hspace{0.4in}-\left( k_{3}-k_{1}\right) _{\nu
}\left( \xi _{20}+\xi _{11}-\xi _{10}\right) ]  \nonumber \\
&&\hspace{1in}\hspace{1in}\hspace{0.1in}+\varepsilon _{\mu \beta \nu \xi
}[\left( k_{1}-k_{2}\right) _{\lambda }\left( \xi _{02}-\xi _{11}-\xi
_{01}\right)  \nonumber \\
&&\hspace{1in}\hspace{1in}\hspace{0.4in}-\left. \left( k_{3}-k_{1}\right)
_{\lambda }\left( \xi _{11}-\xi _{20}+\xi _{10}\right) ]\right\}  \nonumber
\\
&&-\frac{\varepsilon _{\mu \nu \lambda \xi }}{4}\left\{ \left( \left(
k_{3}-k_{1}\right) _{\xi }-\left( k_{1}-k_{2}\right) _{\xi }\right) \left[
I_{\log }-\left( \frac{i}{\left( 4\pi \right) ^{2}}\right) 2\eta _{00}\right]
\right\}  \nonumber \\
&&-\frac{\varepsilon _{\nu \beta \lambda \sigma }}{4}\left(
k_{3}-k_{2}\right) _{\beta }\Delta _{\mu \sigma }-\frac{\varepsilon _{\mu
\beta \lambda \sigma }}{4}\left( k_{3}-k_{2}\right) _{\beta }\Delta _{\nu
\sigma }+  \nonumber \\
&&+\frac{\varepsilon _{\mu \sigma \nu \beta }}{4}\left[ \left(
k_{3}-k_{1}\right) _{\beta }-\left( k_{1}-k_{2}\right) _{\beta }\right]
\Delta _{\lambda \sigma }.
\end{eqnarray}
Putting all results together we write the full expression for the $AVV$
triangle amplitude 
\begin{eqnarray}
T_{\lambda \mu \nu }^{AVV} &=&\left( \frac{i}{\left( 4\pi \right) ^{2}}%
\right) 4\left( k_{3}-k_{1}\right) _{\xi }\left( k_{1}-k_{2}\right) _{\beta
}\left\{ \varepsilon _{\nu \lambda \beta \xi }[\left( k_{3}-k_{1}\right)
_{\mu }\left( \xi _{20}+\xi _{11}-\xi _{10}\right) \right.  \nonumber \\
&&\hspace{1in}\hspace{1in}\hspace{0.4in}\;-\left( k_{1}-k_{2}\right) _{\mu
}\left( \xi _{11}+\xi _{02}-\xi _{01}\right) ]  \nonumber \\
&&\hspace{1in}\hspace{1in}\;\;+\varepsilon _{\mu \lambda \beta \xi }[\left(
k_{3}-k_{1}\right) _{\nu }\left( \xi _{11}+\xi _{20}-\xi _{10}\right) 
\nonumber \\
&&\hspace{1in}\hspace{1in}\hspace{0.4in}\;-\left( k_{1}-k_{2}\right) _{\nu
}\left( \xi _{02}+\xi _{11}-\xi _{01}\right) ]  \nonumber \\
&&\hspace{1in}\hspace{1in}\;\;+\varepsilon _{\mu \nu \beta \xi }[\left(
k_{3}-k_{1}\right) _{\lambda }\left( \xi _{11}-\xi _{20}+\xi _{10}\right) 
\nonumber \\
&&\hspace{1in}\hspace{1in}\hspace{0.4in}\;-\left. \left( k_{1}-k_{2}\right)
_{\lambda }\left( \xi _{02}-\xi _{01}-\xi _{11}\right) ]\right\}  \nonumber
\\
&&-\left( \frac{i}{\left( 4\pi \right) ^{2}}\right) \varepsilon _{\mu \nu
\lambda \beta }\left( k_{3}-k_{1}\right) _{\beta }\left\{ Z_{0}\left( \left(
k_{1}-k_{3}\right) ^{2};m^{2}\right) -Z_{0}\left( \left( k_{2}-k_{3}\right)
^{2};m^{2}\right) \right.  \nonumber \\
&&\hspace{1in}\hspace{1in}+\left[ 2\left( k_{3}-k_{2}\right) ^{2}-\left(
k_{1}-k_{3}\right) ^{2}\right] \xi _{10}+  \nonumber \\
&&\hspace{1in}\hspace{1in}\left. -\;\left( k_{1}-k_{2}\right) ^{2}\xi _{01}+%
\left[ 1-2m^{2}\xi _{00}\right] \right\}  \nonumber \\
&&+\left( \frac{i}{\left( 4\pi \right) ^{2}}\right) \varepsilon _{\mu \nu
\lambda \beta }\left( k_{1}-k_{2}\right) _{\beta }\left\{ Z_{0}\left( \left(
k_{1}-k_{2}\right) ^{2};m^{2}\right) -Z_{0}\left( \left( k_{2}-k_{3}\right)
^{2};m^{2}\right) \right.  \nonumber \\
&&\hspace{1in}\hspace{1in}+\left[ 2\left( k_{3}-k_{2}\right) ^{2}-\left(
k_{1}-k_{2}\right) ^{2}\right] \xi _{01}  \nonumber \\
&&\hspace{1in}\hspace{1in}\left. -\;\left( k_{3}-k_{1}\right) ^{2}\xi _{10}+%
\left[ 1-2m^{2}\xi _{00}\right] \right\}  \nonumber \\
&&+\varepsilon _{\mu \nu \beta \sigma }\left[ \left( k_{3}-k_{1}\right)
_{\beta }-\left( k_{1}-k_{2}\right) _{\beta }\right] \Delta _{\lambda \sigma
}  \nonumber \\
&&+\varepsilon _{\nu \lambda \beta \sigma }\left( k_{3}-k_{2}\right) _{\beta
}\Delta _{\mu \sigma }+\varepsilon _{\mu \lambda \beta \sigma }\left(
k_{3}-k_{2}\right) _{\beta }\Delta _{\nu \sigma }  \nonumber \\
&&+\varepsilon _{\mu \nu \lambda \alpha }\left[ \left( k_{1}+k_{2}\right)
_{\beta }+\left( k_{3}+k_{1}\right) _{\beta }\right] \Delta _{\alpha \beta }.
\end{eqnarray}

The above expression for the $AVV$ triangle is the most general one
concerning the divergences aspects. All the arbitrariness involved is
located in the last four terms. Only the last one is potentially ambiguous.
Following a similar procedure we can also calculate the $T_{\mu \nu }^{PVV}$
amplitude. After taking Dirac traces and using the results (38) and (39) we
get 
\begin{equation}
T_{\mu \nu }^{PVV}=\left( \frac{1}{4\pi ^{2}}\right) m\varepsilon _{\mu \nu
\alpha \beta }\left( k_{1}-k_{2}\right) _{\beta }\left( k_{3}-k_{1}\right)
_{\alpha }\left( \xi _{00}\right) .
\end{equation}

Let us verify the identities (18), (21), and (22) before any assumptions
about the arbitrariness. For the necessary calculations a reasonable
algebraic effort is involved. Substantial simplifications can be achieved if
we note some relations between the structure functions $\xi _{nm}\left(
q,p;m\right) $ and $Z_{k}\left( m^{2};p^{2}\right) .$ They are 
\begin{eqnarray}
\bullet q^{2}\left( \xi _{11}\right) -\left( p\cdot q\right) \left( \xi
_{02}\right) &=&\frac{1}{2}\left\{ -\frac{1}{2}Z_{0}\left(
m^{2};(p+q)^{2}\right) +\frac{1}{2}Z_{0}\left( m^{2};p^{2}\right)
+q^{2}\left( \xi _{01}\right) \right\} \\
\bullet q^{2}\left( \xi _{20}\right) -\left( p\cdot q\right) \left( \xi
_{11}\right) &=&\frac{1}{2}\left\{ -\left[ \frac{1}{2}+m^{2}\xi _{00}\right]
+\frac{p^{2}}{2}\left( \xi _{01}\right) +\frac{3q^{2}}{2}\left( \xi
_{10}\right) \right\} \\
\bullet p^{2}\left( \xi _{02}\right) -\left( p\cdot q\right) \left( \xi
_{11}\right) &=&\frac{1}{2}\left\{ -\left[ \frac{1}{2}+m^{2}\xi _{00}\right]
+\frac{q^{2}}{2}\left( \xi _{10}\right) +\frac{3p^{2}}{2}\left( \xi
_{01}\right) \right\} \\
\bullet p^{2}\left( \xi _{11}\right) -\left( p\cdot q\right) \left( \xi
_{20}\right) &=&\frac{1}{2}\left\{ -\frac{1}{2}Z_{0}\left(
m^{2};(p+q)^{2}\right) +\frac{1}{2}Z_{0}\left( m^{2};q^{2}\right) +p^{2}\xi
_{10}\right\} \\
\bullet q^{2}\left( \xi _{10}\right) -\left( p\cdot q\right) \left( \xi
_{01}\right) &=&\frac{1}{2}\left\{ -Z_{0}\left( m^{2};\left( p+q\right)
^{2}\right) +Z_{0}\left( m^{2};p^{2}\right) +q^{2}\left( \xi _{00}\right)
\right\} \\
\bullet p^{2}\left( \xi _{01}\right) -\left( p\cdot q\right) \left( \xi
_{10}\right) &=&\frac{1}{2}\left\{ -Z_{0}\left( m^{2};\left( p+q\right)
^{2}\right) +Z_{0}\left( m^{2};q^{2}\right) +p^{2}\left( \xi _{00}\right)
\right\} .
\end{eqnarray}
The results so obtained are 
\begin{eqnarray}
\bullet \left( k_{3}-k_{1}\right) _{\mu }T_{\lambda \mu \nu }^{AVV}
&=&\left( k_{3}-k_{1}\right) _{\mu }\Gamma _{\lambda \mu \nu }^{AVV} 
\nonumber \\
&&+\left( \frac{i}{8\pi ^{2}}\right) \varepsilon _{\nu \beta \lambda \xi
}\left( k_{3}-k_{1}\right) _{\xi }\left( k_{1}-k_{2}\right) _{\beta } \\
\bullet \left( k_{1}-k_{2}\right) _{\nu }T_{\lambda \mu \nu }^{AVV}
&=&\left( k_{1}-k_{2}\right) _{\nu }\Gamma _{\lambda \mu \nu }^{AVV} 
\nonumber \\
&&-\left( \frac{i}{8\pi ^{2}}\right) \varepsilon _{\mu \beta \lambda \xi
}\left( k_{3}-k_{1}\right) _{\xi }\left( k_{1}-k_{2}\right) _{\beta } \\
\bullet \left( k_{3}-k_{2}\right) _{\lambda }T_{\lambda \mu \nu }^{AVV}
&=&\left( k_{3}-k_{2}\right) _{\lambda }\Gamma _{\lambda \mu \nu }^{AVV} 
\nonumber \\
&&+\left( \frac{i}{4\pi ^{2}}\right) \varepsilon _{\mu \xi \nu \beta }\left(
k_{3}-k_{1}\right) _{\xi }\left( k_{1}-k_{2}\right) _{\beta }\left[
2m^{2}\xi _{00}\right] ,
\end{eqnarray}
where we have defined 
\begin{eqnarray}
\Gamma _{\lambda \mu \nu }^{AVV} &=&\varepsilon _{\mu \nu \beta \xi }\left[
\left( k_{2}-k_{1}\right) _{\beta }+\left( k_{3}-k_{1}\right) _{\beta }%
\right] \Delta _{\lambda \xi }  \nonumber \\
&&+\varepsilon _{\nu \lambda \beta \xi }\left( k_{3}-k_{2}\right) _{\beta
}\Delta _{\mu \xi }+\varepsilon _{\mu \lambda \beta \xi }\left(
k_{3}-k_{2}\right) _{\beta }\Delta _{\nu \xi }  \nonumber \\
&&+\varepsilon _{\mu \nu \lambda \alpha }\left[ \left( k_{1}+k_{2}\right)
_{\beta }+\left( k_{3}+k_{1}\right) _{\beta }\right] \Delta _{\alpha \beta }.
\end{eqnarray}
The above expressions can be rewritten in terms of other structures if we
observe the equation (81) and (44). As an example, note that 
\begin{equation}
\left( k_{3}-k_{2}\right) _{\lambda }\Gamma _{\lambda \mu \nu
}^{AVV}=2\varepsilon _{\mu \nu \alpha \beta }\left[ (k_{1}-k_{3})_{\beta
}(k_{1}+k_{3})_{\xi }+(k_{2}-k_{1})_{\beta }(k_{1}+k_{2})_{\xi }\right] \
\triangle _{\xi \alpha },
\end{equation}
which means that 
\begin{equation}
\left( k_{3}-k_{2}\right) _{\lambda }\Gamma _{\lambda \mu \nu }^{AVV}=T_{{%
\mu }{\nu }}^{AV}(k_{1},k_{2};m)-T_{{\nu \mu }}^{AV}(k_{3},k_{1};m),
\end{equation}
and in consequence the equation (90) becomes identical to the equation (18).

When we compare the results (88)-(90) with the expected identities, which
are the equations (18), (21), and (22), it is immediate to see that the
first one is satisfied but the two remaining ones present a disagreement. As
a consequence the respective Ward identities, obtained by the addition of
the crossed diagrams, will be violated.

Now if the consistent value for the $AV$ amplitude is assumed, i.e.,
adopting $\triangle _{\xi \alpha }=0$, the equations (88)-(90) become 
\begin{eqnarray}
\bullet \left( k_{3}-k_{1}\right) _{\mu }T_{\lambda \mu \nu }^{AVV}
&=&\left( \frac{i}{8\pi ^{2}}\right) \varepsilon _{\nu \beta \lambda \xi
}\left( k_{3}-k_{1}\right) _{\xi }\left( k_{1}-k_{2}\right) _{\beta } \\
\bullet \left( k_{1}-k_{2}\right) _{\nu }T_{\lambda \mu \nu }^{AVV}
&=&-\left( \frac{i}{8\pi ^{2}}\right) \varepsilon _{\mu \beta \lambda \xi
}\left( k_{3}-k_{1}\right) _{\xi }\left( k_{1}-k_{2}\right) _{\beta } \\
\bullet \left( k_{3}-k_{2}\right) _{\lambda }T_{\lambda \mu \nu }^{AVV}
&=&-2mi\left\{ T_{\mu \nu }^{PVV}\right\} .
\end{eqnarray}
The contribution of the crossed diagram gives us 
\begin{eqnarray}
\bullet \left( l_{1}-l_{2}\right) _{\mu }T_{\lambda \nu \mu }^{AVV}
&=&-\left( \frac{i}{8\pi ^{2}}\right) \varepsilon _{\nu \beta \lambda \xi
}\left( l_{3}-l_{1}\right) _{\xi }\left( l_{1}-l_{2}\right) _{\beta } \\
\bullet \left( l_{3}-l_{1}\right) _{\nu }T_{\lambda \nu \mu }^{AVV}
&=&\left( \frac{i}{8\pi ^{2}}\right) \varepsilon _{\mu \beta \lambda \xi
}\left( l_{3}-l_{1}\right) _{\xi }\left( l_{1}-l_{2}\right) _{\beta } \\
\bullet \left( l_{3}-l_{2}\right) _{\lambda }T_{\lambda \nu \mu }^{AVV}
&=&-2mi\left\{ T_{\nu \mu }^{PVV}\right\} ,
\end{eqnarray}
so that we get 
\begin{eqnarray}
\bullet p_{\mu }T_{\lambda \mu \nu }^{A\rightarrow VV} &=&\left( \frac{i}{%
4\pi ^{2}}\right) \varepsilon _{\nu \beta \lambda \xi }p_{\xi }p_{\beta
}^{\prime } \\
\bullet p_{\nu }^{\prime }T_{\lambda \mu \nu }^{A\rightarrow VV} &=&-\left( 
\frac{i}{4\pi ^{2}}\right) \varepsilon _{\mu \beta \lambda \xi }p_{\xi
}p_{\beta }^{\prime } \\
\bullet q_{\lambda }T_{\lambda \mu \nu }^{A\rightarrow VV} &=&-2mi\left\{
T_{\mu \nu }^{P\rightarrow VV}\right\} .
\end{eqnarray}

These above results are perfectly compatible with what states the
Sutherland-Veltman paradox: two Ward identities and the low-energy theorem
have been obtained violated. Only the axial vector symmetry relation is
satisfied. At this point the most important aspect is the fact that, in
order to verify the violations, no fundamental symmetries were broken. The
anomalies are not conditioned or associated to the typical arbitrariness
present in the perturbative calculations. No particular treatment has been
used. The violations have emerged in a natural way within a general
calculational strategy that treats all the physical amplitudes of all
theories and models in the same way. As a result of this treatment, i.e., by
using the same expressions for the integrals we have used here, it can be
shown that in all other three-point functions the symmetries are preserved
(except obviously for the $AAA$ which is anomalous too \cite{Orimar-TESE}).
In spite of such verification has been in fact performed \cite{Orimar-TESE}
we can convince ourselves by a simple argument. All the results of the DR
can be immediately mapped by the choices $\Delta =\nabla =\Box =0$. Then if
the DR is accepted as a consistent treatment, concerning the avoidance of
ambiguities and symmetry preservation, it is almost immaterial such explicit
verification. After these discussions we can return to the pion decay
problem.

\section{Pion Decay and Anomalies; Final Remarks and Conclusions}

Given the obtained results in the previous section and considering all the
argumentation put in the remaining ones, what is the situation for the pion
decay relative to the $AVV$ triangle anomaly? From general aspects, this
specific phenomenology is only a particular consequence of a theory or model
where the involved final states, the pseudo-scalar and the photon fields,
can be connected through the coupling to the quark-antiquark fields.
However, in a theory or model where we also have an axial-vector, there are
two different paths to construct the amplitude which must describe the pion
decay in two photons. The first is the direct evaluation of the $PVV$
amplitude, equation (81), which in fact produces a low-energy limit
compatible with the pion decay rate. The second is to evaluate the $AVV$
amplitude and after this the $PVV$ one through the identity (see equation
(18)) 
\begin{equation}
T_{\mu \nu }^{PVV}=\frac{i}{2m}\left\{ \left( k_{3}-k_{2}\right) _{\lambda
}T_{\lambda \mu \nu }^{AVV}+T_{\nu \mu }^{AV}\left( k_{3},k_{1};m\right)
-T_{\mu \nu }^{AV}\left( k_{1},k_{2};m\right) \right\} .
\end{equation}
The same $\ AVV$ amplitude must also satisfy (see equation (21) and (22)) 
\begin{eqnarray}
\bullet \left( k_{3}-k_{1}\right) _{\mu }T_{{\lambda }{\mu }{\nu }}^{AVV}
&=&T_{{\lambda \nu }}^{AV}(k_{1},k_{2};m)-T_{{\lambda \nu }}^{AV}\left(
k_{3},k_{2};m\right) \\
\bullet \left( k_{1}-k_{2}\right) _{\nu }T_{{\lambda }{\mu }{\nu }}^{AVV}
&=&T_{{\lambda \mu }}^{AV}(k_{3},k_{2};m)-T_{{\lambda \mu }%
}^{AV}(k_{3},k_{1};m),
\end{eqnarray}
and then the things do not seem to work as expected.

This is due to the fact that by using the $PCAC$ hypothesis, the $LSZ$
formalism and the standard procedures of the current algebra, it is possible
to derive a low-energy theorem which states that 
\begin{equation}
\lim_{q_{\lambda }\rightarrow 0}q_{\lambda }T_{\lambda \mu \nu
}^{A\rightarrow VV}=0,
\end{equation}
and in addition from the current algebra methods 
\begin{eqnarray}
\bullet p_{\mu }T_{\lambda \mu \nu }^{A\rightarrow VV} &=&p_{\nu }^{\prime
}T_{\lambda \mu \nu }^{A\rightarrow VV}=0 \\
\bullet q_{\lambda }T_{\lambda \mu \nu }^{A\rightarrow VV} &=&-2miT_{\mu \nu
}^{P\rightarrow VV},
\end{eqnarray}
which creates a paradoxical situation involving the pion decay. If the
low-energy limit (106) is satisfied by the $AVV$ amplitude, due to the axial
Ward identity (108), the $PVV$ amplitude should also vanish at this limit.
Consequently, the pion decay rate is not obtained as the experimental one.
So, if the $AVV$ and $PVV$ satisfy the four equations above, the pion does
not decay. What can we make in order to reconciliate the theory with the
phenomenology? There are two different options:

{\bf i) The traditional approach:}

The usual procedure is the admittance of an anomalous term on the right hand
side of the equation (108) which must have precisely the value that
corresponds to the one necessary for the correct pion decay rate. In order
to justify the presence of such term, it is looked at the right hand side of
the identity (103) and it is attributed to the $AV$ amplitude the value 
\begin{equation}
T_{{\mu \nu }}^{AV}(k_{i},k_{j};m)=i\varepsilon _{\mu \nu \alpha \beta
}\left( k_{i}-k_{j}\right) _{\alpha }\left( k_{j}+k_{i}\right) _{\beta
}\left( \frac{1}{32\pi ^{2}}\right) ,
\end{equation}
where $k_{i}-k_{j}$ are the external momenta. The combination $k_{j}+k_{i}$
are undefined quantities. After the substitution of this expression in the
equations (103), (104), and (105) the values for the undefined quantities
are chosen in order to get the desirable situation. The justification for
the existence of an anomaly in the problem resides in the fact that it is
not possible to obtain the three summations of $AV$ amplitudes involved in
the equations (103), (104), and (105) simultaneously vanishing. Given this
impossibility, which implies in the violation of at least one Ward identity,
we are authorized to perform the most adequate choice for the undefined
quantities $k_{1}+k_{2}$,$\;k_{3}+k_{2}$ and $k_{1}+k_{3}$. The two vector
Ward identities are chosen to be preserved and the axial one is assumed
violated precisely by the amount necessary to the low-energy limit required
by the pion decay phenomenology.

{\bf ii) The proposed interpretation:}

In view of the argumentations we have presented in this contribution, the
above described interpretation for the situation involving the pion decay
phenomenology, in spite of being well succeeded in the furnishment of a
justification for the needed anomalous term, is founded in ingredients which
can be questionable. The main point is the attribution of a nonzero value
for the $AV$ two-point function amplitude. We have argued that this means to
violate the unitarity, CPT and the axial Ward identity, once 
\begin{equation}
\left( k_{j}-k_{i}\right) _{\mu }T_{\mu \nu }^{AV}=0,
\end{equation}
if it is non-vanishing. This procedure makes use of the potentially
ambiguous character of perturbative physical amplitudes. This means to
assume that we cannot make definite predictions within the perturbative
solutions of QFT in general. Only adjustments for well-known
phenomenologies, like that of the pion decay, can be obtained due to the
fact that, in all the calculations, completely undefined pieces can be
present in a calculated amplitude. The justification through the infinities
or ambiguities is also in contradiction with the logical expectations: the
anomaly involved in the pion decay through the $AVV$ amplitude is predicted
for the exact physical amplitudes, in which case we certainly do not have
infinities and ambiguities. So it does not seem to be reasonable any
justifications, even for perturbative evaluations of the involved
amplitudes, based on ingredients which are exclusive of perturbative
calculations. In addition, now specifically referring to the perturbative
aspects, the adopted procedure is in contradiction with the interpretations
for the divergences included in the DR technique, which represents our best
reference concerning the consistency in perturbative calculations. The
interpretations of the traditional approach, to justify the $AVV$ anomaly,
cannot be mapped in the ones of DR, in a problem where both procedures can
be applied \cite{Orimar-Orildo,Orimar-PRD2}.

In the proposed interpretation, which maps the DR results in all places
where it is possible to apply this technique, the properties for the
divergent integrals are taken as universal ones. The $AV$ two-point function
is obtained identically zero and the ambiguities are automatically
eliminated. The Ward identities for the $AVV$ amplitude are verified over
the explicit expression. The $AV$ structures, which are expected to be
present in contracted expressions, are in fact identified but the anomalous
term, present in the calculation, is not associated to such potentially
ambiguous structures. The violating term emerges as a particular property of
the $AVV$ amplitude, which is independent of the infinities or ambiguities
involved, as should be expected. All other two- and three-point functions,
which are constructed by the same set of Feynman integrals, have their Ward
identities preserved (mapping with DR) \cite{Orimar-TESE,Orimar-Orildo}.

After these important remarks we can conclude our proposed interpretation.
In the construction of theories where this amplitude is present, we have no
option than to choose, in an arbitrary way, what is the symmetry content we
want to have. All the options are in principle equivalent, due to the
fundamental nature of the anomalies. If we want to have a theory consistent
with the pion decay phenomenology, which seems to be the most important
ingredient, we need to redefine the $AVV$ amplitude in order to impose the
correct behavior to the limit of the vanishing axial vertex momentum 
\begin{equation}
\left( T_{\lambda \mu \nu }^{A\rightarrow VV}(p,p^{\prime })\right)
_{phy}=T_{\lambda \mu \nu }^{A\rightarrow VV}(p,p^{\prime })-T_{\lambda \mu
\nu }^{A\rightarrow VV}\left( 0\right) ,
\end{equation}
where 
\begin{equation}
T_{\lambda \mu \nu }^{A\rightarrow VV}(0)=-\left( \frac{i}{4\pi ^{2}}\right)
\varepsilon _{\mu \nu \lambda \xi }\left[ p_{\xi }-p_{\xi }^{\prime }\right]
,
\end{equation}
is, as it should be, the required anomalous term. Consequently, we get for
the $AVV$ physical amplitude 
\begin{eqnarray}
\bullet p_{\nu }^{\prime }\left( T_{\lambda \mu \nu }^{A\rightarrow
VV}\right) _{phy} &=&0 \\
\bullet p_{\mu }\left( T_{\lambda \mu \nu }^{A\rightarrow VV}\right) _{phy}
&=&0 \\
\bullet q_{\lambda }\left( T_{\lambda \mu \nu }^{A\rightarrow VV}\right)
_{phy} &=&-2mi\left\{ T_{\mu \nu }^{P\rightarrow VV}\right\} -\left( \frac{i%
}{2\pi ^{2}}\right) \varepsilon _{\mu \nu \alpha \beta }p_{\alpha }p_{\beta
}^{\prime }.
\end{eqnarray}

There is certainly an arbitrariness in the imposition of such redefinition,
but it is intrinsic to the problem and is in agreement with the
renormalization procedures. The degree of arbitrariness certainly imposes a
price to be paid that is more acceptable because it does not plague other
amplitudes or problems. Anyway, in the traditional approach to justify the
anomaly, an arbitrary choice is always necessary at the end. The difference
is that such a choice is taken over an undefined quantity. Here the physical
amplitudes do not become ambiguous quantities and the most fundamental
space-time symmetry, that is the space-time homogeneity, need not be assumed
broken neither CPT or unitarity. The redefined amplitude, allows the neutral
pion decay in agreement with the experimental data, as it should be.

{\bf Acknowledgements}: G. Dallabona acknowledges a grant from CNPq/Brazil.

\end{document}